\documentclass[sigconf,screen]{acmart}
\usepackage{balance}
\usepackage{xspace}
\usepackage{enumitem}
\usepackage{algorithmic}
\usepackage[ruled,lined,linesnumbered,vlined,algo2e]{algorithm2e}
\definecolor{codegray}{rgb}{0.5,0.5,0.5}

\SetCommentSty{mycommfont}
\usepackage{subfigure}
\usepackage{cleveref}
\usepackage{multirow}
\usepackage{makecell}
\usepackage{array} 
\usepackage{microtype}
\usepackage{color}
\usepackage{xcolor}
\definecolor{springgreen}{RGB}{0,205,102}
\usepackage{colortbl}
\usepackage{flushend}
\usepackage[utf8]{inputenc}
\usepackage[T1]{fontenc}

\newcommand{\sihan}[1][\textcolor{purple}]{#1}

\newcommand{\revise}[1]{\textcolor{black}{#1}}

\newcommand{\tool}{\textsf{LiResolver}\xspace}

\AtBeginDocument{%
  \providecommand\BibTeX{{%
    \normalfont B\kern-0.5em{\scshape i\kern-0.25em b}\kern-0.8em\TeX}}}

\usepackage[font=normalsize,skip=2pt]{caption}
\newcommand{\distance}{2pt}
\setcopyright{acmlicensed}
\acmPrice{15.00}
\acmDOI{10.1145/3597926.3598085}
\acmYear{2023}
\copyrightyear{2023}
\acmSubmissionID{issta23main-p165-p}
\acmISBN{979-8-4007-0221-1/23/07}
\acmConference[ISSTA '23]{Proceedings of the 32nd ACM SIGSOFT International Symposium on Software Testing and Analysis}{July 17--21, 2023}{Seattle, WA, USA}
\acmBooktitle{Proceedings of the 32nd ACM SIGSOFT International Symposium on Software Testing and Analysis (ISSTA '23), July 17--21, 2023, Seattle, WA, USA}
\received{2023-02-16}
\received[accepted]{2023-05-03}

\begin{document}

\title{\tool: License Incompatibility Resolution for Open Source Software}
%\title{\tool: Resolving License Incompatibility Issues for \\Open Source Software}

%\author{Sihan Xu, Ya Gao, Lingling Fan\textsuperscript{$\ast$}, Linyu Li, Xiangrui Cai, and Zheli Liu}

\author{Sihan Xu}
\email{xusihan@nankai.edu.cn}
\affiliation{%
  \institution{DISSec, NDST, College of Cyber Science, Nankai University}
  \city{Tianjin}
  \country{China}
}

\author{Ya Gao}
\email{gaoya_cs@mail.nankai.edu.cn}
\affiliation{
  \institution{DISSec, NDST, College of Cyber Science, Nankai University}
  \city{Tianjin}
  \country{China}
}

\author{Lingling Fan}
\email{linglingfan@nankai.edu.cn}
\affiliation{
  \institution{DISSec, NDST, College of Cyber Science, Nankai University}
  \city{Tianjin}
  \country{China}
}
\authornote{Lingling Fan is the corresponding author. Email: linglingfan@nankai.edu.cn}

\author{Linyu Li}
\email{linyuli@mail.nankai.edu.cn}
\affiliation{
  \institution{DISSec, NDST, College of Cyber Science, Nankai University}
  \city{Tianjin}
  \country{China}
}

\author{Xiangrui Cai}
\email{caixr@nankai.edu.cn}
\affiliation{
  \institution{DISSec, NDST, College of Computer Science, Nankai University}
  \city{Tianjin}
  \country{China}
}

\author{Zheli Liu}
\email{liuzheli@nankai.edu.cn}
\affiliation{
  \institution{DISSec, NDST, College of Cyber Science, Nankai University}
  \city{Tianjin}
  \country{China}
}

\renewcommand{\shortauthors}{Sihan Xu, Ya Gao, Lingling Fan, Linyu Li, Xiangrui Cai, and Zheli Liu}

\begin{abstract}
Open source software (OSS) licenses regulate the conditions under which OSS can be legally reused, distributed, and modified. However, a common issue arises when incorporating third-party OSS accompanied with licenses, i.e., license incompatibility, which occurs when multiple licenses exist in one project and there are conflicts between them. Despite being problematic, fixing license incompatibility issues requires substantial efforts
due to the lack of license understanding and complex package dependency.
In this paper, we propose \tool, a fine-grained, scalable, and flexible tool to resolve license incompatibility issues {for open source software}. Specifically, it first understands the semantics of licenses through fine-grained entity extraction and relation extraction. Then, it detects and resolves license incompatibility issues by recommending official licenses in priority. When no official licenses can satisfy the constraints, it generates a custom license as an alternative solution. Comprehensive experiments demonstrate the effectiveness of \tool, with 4.09\% \revise{false positive (FP)} rate and 0.02\% \revise{false negative (FN)} rate for incompatibility issue localization, and 62.61\% of 230 real-world incompatible projects resolved by \tool. %Furthermore, we also evaluate the impacts of license hierarchy and copyright holder detection on the effectiveness of license incompatibility resolution. 
We discuss the feedback from OSS developers and the lessons learned from this work. All the datasets and the replication package of \tool have been made publicly available to facilitate follow-up research.
\end{abstract}




\maketitle

\section{Introduction}
% the definition of OSS and OSS license
Open source software (OSS) significantly facilitates software development. Developers do not need to reinvent the wheel but only focus on their unique workflows and features. Despite the benefits, misuse of OSS might also induce security issues and legal risks~\cite{licensing2004software,infringement-case,chinacase,infringement-case2,liu2022demystifying}. Many issues arise from OSS licenses, which regulate the conditions under which OSS can be \textit{legally} reused, distributed, and modified. Generally, an OSS license plays the role as a contract signed between the software owner and the user who reuses the code or binary. Once a user of OSS does not conform to the conditions dictated by OSS licenses, license violation occurs and legal risks are induced such as copyright infringement~\cite{infringement-case3}. 

OSS licenses, which are legally binding, protect both the rights of software owners and the freedom of OSS.
Nevertheless, a common issue arises when incorporating third-party OSS accompanied with licenses, i.e., \textit{license incompatibility}, which occurs when multiple OSS licenses exist in the same project and there are conflicts between the conditions they state~\cite{gangadharan2012managing,jss`17-Kapitsaki-SPDX}. For example, ``CANNOT sublicense'' declared by the license of a third-party component but ``CAN sublicense'' declared by the license of the whole project {are incompatible with each other}. In this case, conforming to the project license cannot ensure that the requirements declared by the license of the third-party component are satisfied simultaneously. License incompatibility can also occur among the licenses of multiple third-party components, 
%For example, 
{e.g.,} ``MUST contact author'' from the license of one component but ``CANNOT contact author'' from the license of another component. In this case, at least one license is being violated due to the license incompatibility issue.

License incompatibility represents a serious threat to all stakeholders from a legal perspective. An empirical study on 1,846 GitHub projects shows that 72.91\% of the investigated projects suffered from license incompatibility~\cite{xu2021lidetector}. Despite being problematic, fixing license incompatibility issues requires substantial efforts to either replace the reused code or migrate licenses. Worse still, developers struggle when multiple licenses are involved in the same project~\cite{almeida2017software}. To address this issue, Kapitsaki et al.~\cite{jss`17-Kapitsaki-SPDX,kapitsaki2019modeling} proposed a solution based on the Software Package Data Exchange (SPDX) specification~\cite{spdx}. They designed a directed graph with the compatibility relationships between licenses, based on which
%Based on the graph,
they proposed SPDX Violation Tools (SPDX-VT) to detect license incompatibility and recommend licenses to handle incompatibility issues. Besides, Liu et al.~\cite{liu2019predicting} also proposed Automatic License Prediction (ALP) to prevent license incompatibility issues induced by software changes. 

Despite the progress, there are mainly three problems that limit the application of previous studies. \textbf{First}, previous studies only cover a small predefined set of licenses (e.g., 20 licenses supported by SPDX-VT~\cite{kapitsaki2019modeling} and 25 licenses supported by ALP~\cite{liu2019predicting}). However, there are 489 licenses and versions in SPDX~\cite{spdx}, not to mention potentially a considerable %amount
{number} of license exceptions and custom licenses. \textbf{Second}, previous works only considered the names and versions of some well-known licenses, lacking the capability to comprehend license texts and understand the inherent reasons of license incompatibility, which limits their abilities to resolve incompatibility issues in a flexible and fine-grained way. % when licenses are changed, updated, or customized by developers. 
\textbf{Third}, the solutions to license incompatibility have been mainly focused on recommending an official license as the new license for the whole project (i.e., the project license). Whereas, %our preliminary study on \sihan{XX} projects shows that only  \sihan{XX}\% incompatibility issues can be fixed by recommending a new license from these candidates. 
this paper observes that %many license incompatibility issues can not be resolved only by official licenses. Moreover, 
there may exist other licenses whose copyright holders are the same with the project license, and thus can also be changed by the licensor for resolving license incompatibility issues. 

\begin{figure*}[]
    \centering
    \includegraphics[width=0.95\textwidth]{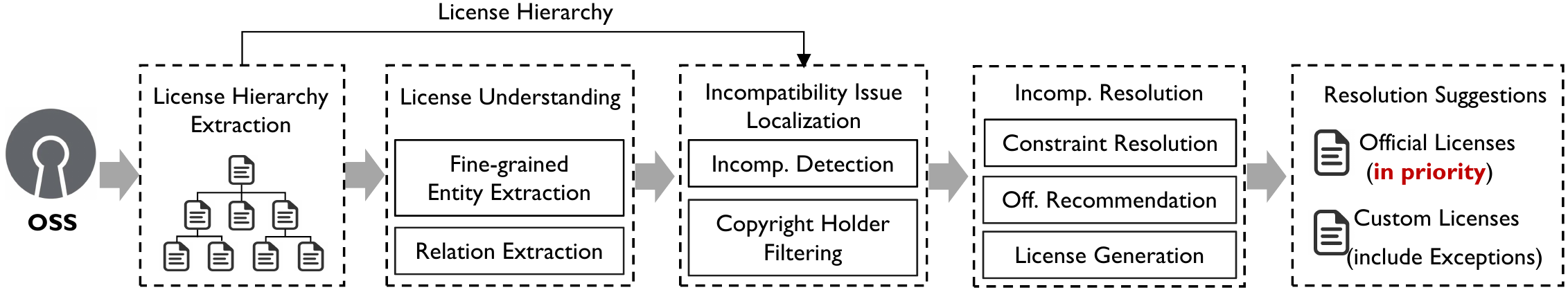}
    %\vspace{-2mm}
    \caption{Overview of \tool}
    \label{fig_overview}
\end{figure*}

To address the aforementioned issues, it is desirable to provide a fine-grained, scalable, and flexible solution to resolve license incompatibility issues.
To this end, we propose \tool, an %automatic
{automated} tool to fix license incompatibility issues for open source software. Specifically, given an OSS, it first extracts all licenses accompanied with the project, and organizes them with a hierarchical structure that represents the licensing scope. Then, it provides a fine-grained understanding of the regulations stated by each license, based on which \tool detects and localizes license incompatibility issues. Finally, it resolves the constraints hidden in the licensing context, and provides flexible suggestions for users to choose from. Comprehensive experiments on real-world projects demonstrate the effectiveness of \tool, with {4.09\%} \revise{false positive (FP)} rate and {0.02\%} \revise{false negative (FN)} rate for incompatibility issue localization, and {62.61\%} success rate for resolving 230 incompatible OSS on GitHub. %Furthermore, we also evaluate the impacts of license hierarchy, copyright holder, as well as license exceptions on the effectiveness of incompatibility resolution. 
%The results exhibit the importance of license incompatibility resolution and the practicability of \tool. 
%Finally, we discuss the feedback from OSS developers, the lessons learned from this work, as well as the limitations and threats to validity. 
We made all datasets and the replication package publicly available to facilitate follow-up research.

In summary, we made the following novel contributions:
\begin{itemize}
    \item We propose \tool, a fine-grained, scalable, and flexible approach to automatically interpret licenses, localize license incompatibility issues, and provide licensors with useful suggestions for resolving issues. %and made all datasets and source code publicly available. %which provides software owners with useful suggestions to reduce the legal risks induced by incorporating third-party packages.
    \item Comprehensive experiments demonstrate the effectiveness of \tool, with 4.09\% FP rate and 0.02\% FN rate for incompatibility issue localization, and 62.61\% of 230 real-world incompatible projects resolved by \tool.  
    \item We further investigate the impacts of license hierarchy,
copyright holders, and license exceptions on the effectiveness
of license incompatibility resolution. We made all the datasets and the replication package publicly available to facilitate follow-up research~\cite{ourwebsite}. 
%on Github \ling{https://github.com/anonymous123rainy/LiResolver} 
\end{itemize}
% previous studies about how to identify and fix such issues

% the motivation and the methodology of this paper

% the contribution of this paper

%\input{background.tex}
\section{Approach}

\subsection{Overview}

This section details our tool, \tool, a hybrid approach that automatically understands license texts, detects license incompatibility, and provides useful suggestions for license incompatibility resolution.
\revise{The input of \tool is the path of a folder where the target OSS is stored.} As shown in Figure~\ref{fig_overview}, given an OSS, \tool first extracts all involved licenses for further analysis, including the licenses accompanied with integrated software packages. After preprocessing the extracted licenses, the main components of \tool include three parts: \textbf{(1)} License understanding. As a fundamental step, \tool first understands the regulations implied by licenses by fine-grained entity extraction and relation extraction, so as to provide detailed information for further incompatibility analysis and resolution. \textbf{(2)} Incompatibility issue localization. Based on the fined-grained understandings of licenses, \tool detects and localizes license incompatibility issues by considering both the hierarchical structure and copyright holders of licenses. \textbf{(3)} License incompatibility resolution. Given a license involved in an incompatibility issue, \tool fixes the issue by resolving the constraints obtained from the parent and child licenses of the target license. Finally, \tool gives suggestions for license incompatibility resolution by recommending official licenses in priority. When none of existing licenses are compatible with the license context, \tool generates a custom license as an alternative solution. \revise{\tool stores the license incompatibility issues, their locations, the corresponding suggestions, i.e., the name of recommended official licenses or the full-text of custom licenses in an output file.} 

\subsection{{License Hierarchy Extraction}}\label{sec:hie}

\begin{figure}

	\centering
	\subfigure[License Compatibility Analysis without Hierarchy Extraction]{
		\begin{minipage}{7cm} 
                        \includegraphics[width=\textwidth]{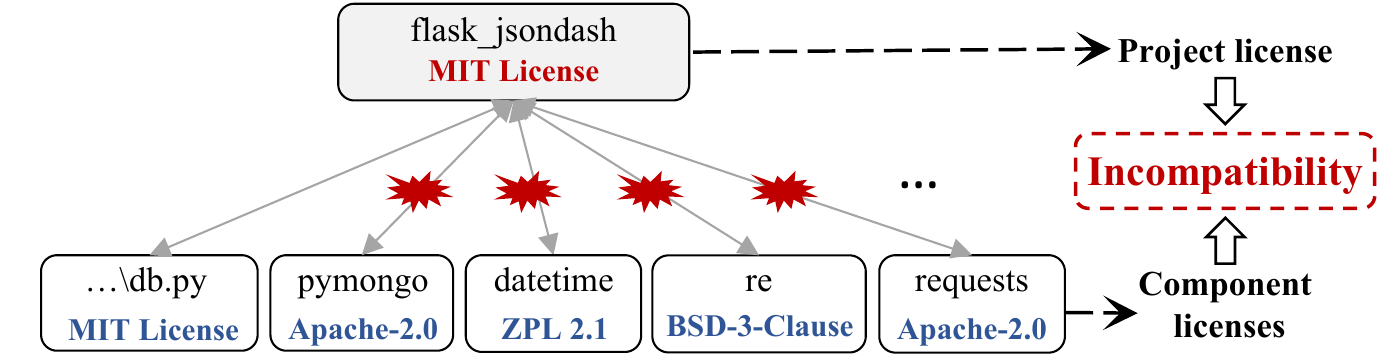} \\

		\end{minipage}\label{withoutHie}
	}

	\subfigure[License Compatibility Analysis with Hierarchy Extraction]{
		\begin{minipage}{7cm}%[b]%{0.2\textwidth}
			\includegraphics[width=\textwidth]{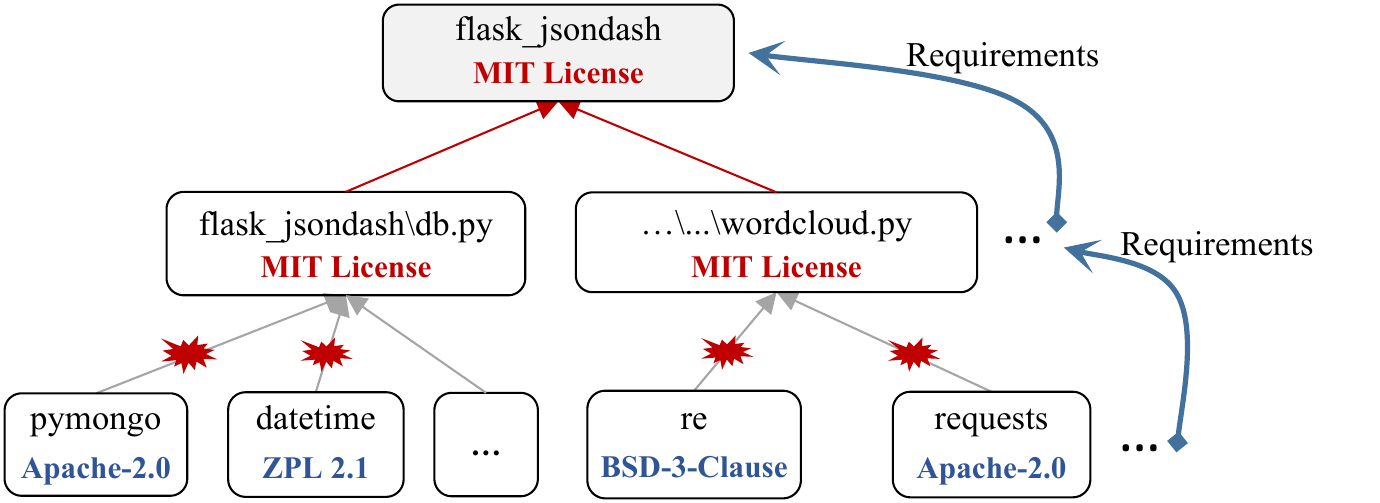} \\
		\end{minipage}\label{withHie}
	}
 \caption{Illustration of License Hierarchy} 
 \label{hierarchy}
\end{figure}

% Given the file hierarchical structure of open source software project explicitly, its 
OSS, especially in the form of frameworks or libraries, is often used as a component of software products, which can be recursively integrated into larger software. 
The hierarchical structure of OSS accompanied with licenses leads to the hierarchical structure of licenses. 
\revise{Previous studies regarded all licenses in a project independently and equally~\cite{jss`17-Kapitsaki-SPDX,kapitsaki2019modeling,xu2021lidetector}, regardless of their licensing scopes.} For instance, \Cref{hierarchy} illustrates a real-world example from Flask JSONDash~\cite{url-example1}, {a popular OSS to create chart dashboards}. In this example, the project license, which states the regulations for the whole project, is the MIT license in the main directory of the project. As shown in \Cref{withoutHie}, there are four component licenses in the modules or integrated third-party packages which are incompatible with the project license. The reason is that the component licenses are more restrictive than the project license, so that anyone who conforms to the project license (i.e., the MIT license) may violate the component licenses (e.g., Apache-2.0). To address this issue, one should either replace the third-party software packages or change the project license in the root directory.
\revise{In this paper, we observe that the location of a license determines its scope of licensing, which affects license incompatibility detection and resolution.} \Cref{withHie} depicts a part of the license hierarchy of JSONDash~\cite{url-example1}. It can be seen that \textit{pymongo} and \textit{datetime} are two third-party packages imported by the source code file \textit{db.py}. However, their licenses (i.e., Apache-2.0 and ZPL-2.1) are incompatible with the license of \textit{db.py} (i.e., the MIT license). In this case, even if the project owner changes the project license, license incompatibility still remains. %License incompatibility may occur anywhere in the project.

To address this issue, we extract the hierarchical structure of all involved licenses in a project. Specifically, \tool first builds the hierarchical structure according to the file structure. Then, it extracts three types of licenses from the project, i.e., the declared licenses, the referenced licenses, and the inline licenses. The \textit{declared licenses} state rights and obligations explicitly in license files (e.g., LICENSE). The \textit{referenced licenses} are obtained by their names, versions, direct links, or the websites of incorporated third-party packages. The \textit{inline licenses} typically appear at the beginning of source code files. Different licenses may have different licensing scopes. For example, an inline license is only responsible for the source code within the same source code file. The scope of the license accompanied with a third-party software package is only the package itself. \revise{After obtaining the original file structure, \tool removes the nodes without licenses from bottom to up and links its children nodes to its parent node if necessary.} %\tool merges the nodes without licenses and obtains the license hierarchy as shown in \Cref{withHie}. 
Finally, the licensing scopes of all licenses can be obtained by the hierarchy as shown in \Cref{withHie}, as well as the dependencies between them, which can be further used for license incompatibility resolution.

%which is necessary because, only file structure may lost some significant information (e.g., the implicit licenses of imported packages in code files), 
% the licenses of imported packages in code files take equal importance with other obvious licenses in texts
%or display some unnecessary information (e.g., if some modules with their license absent, their child modules should act a direct influence on the license compliance of higher modules). 

%In the example of ~\ref{XX}, ...., 

\subsection{License Understanding}\label{sec:und}

After obtaining the license hierarchy, \tool \textit{automatically} understands the semantics of each license as a fundamental step. %However, license texts are typically long and complicated, which makes it difficult to interpret them. The problem is exacerbated by the large number of licenses, variants, and custom licenses. 
\revise{There are mainly two reasons for automating license understanding. (1) There exist a variety of official licenses (i.e., 552), and the number keeps growing~\cite{url-spdx-github}. However, manually extracting the detailed information from long and complicated licenses is time-consuming and labor-intensive. (2) A previous study investigated 1,846 projects and obtained 5,777 unique licenses, among which 24.56\% were custom licenses with flexible expressions~\cite{xu2021lidetector}. Motivated by them, it is valuable to automatically understand licenses to facilitate further analysis.}
In this paper, we cast license understanding as an \textbf{information extraction} problem, which can be further decomposed into an entity extraction task %(i.e., the extraction of the \textit{action}, \textit{object}, \textit{attitude},  and \textit{condition} of a license term) 
and a relation extraction task. %(i.e., the classification of the relations between these entities like \textit{action-object}).

%Hence, the effort on automated license \sihan{interpretation?} does not only facilitate license incompatibility detection, but also provides guidance for repairing license incompatibility issues. % its essentials rapidly, but 
% help detect license incompatibility more accurately and 
%help generate repair texts with more lawful compliance. 

\subsubsection{{Fine-grained Entity Extraction}}
%\ling{may briefly mention why we need such fine-grained entity? the necessity. such as help to further repair.}
%To provide detailed information, we decompose each license into a set of terms which describe the conditions of software use. 
\revise{Unlike previous works that regarded each right/obligation as a single entity~\cite{xu2021lidetector}, in this paper, we observe that each right/obligation could be decomposed into four types of
entities (i.e., \textit{action}, \textit{object}, \textit{attitude}, and \textit{condition}) and construct a fine-grained structure to model it. Each entity, regardless of its type, is a flexible expression that describes a specific action, object, attitude, or condition in the form of a word, a phrase, or even a sentence in a license. \revise{For instance, from ``\textit{you are allowed to distribute modified works}'', the previous study~\cite{xu2021lidetector} identified the term ``Distribute'' and inferred an attitude CAN. However, \tool extracts the action (``distribute''), the object (``modified works''), the attitude (``are allowed to''), and the condition if there exists.} The definitions of four types of entities are as follows.}
%Then, we further model each license term with a fine-grained structure by four types of entities described as follows. 

\textbf{Action}: the process of %excising 
{exercising} a granted right or performing an obligation enforced by a license, e.g., \textit{Disclose Source}. 

\textbf{Object}: the object to be acted upon, which is typically a software artifact such as source code or binary.

%\noindent\textbf{Performer}: the person or organization that grants the rights or conveys the obligations, e.g., \textit{GPL corporation} in GPL 2.0.
\textbf{Attitude}: the attitudes of licensors towards licensees to grant or reserve rights, or enforce obligations.

\textbf{Condition}: the obligations that licensees must comply with to exercise the granted rights.

%\ling{revise.}

\begin{figure}[t]
    \centering
    \includegraphics[width=0.9\linewidth]{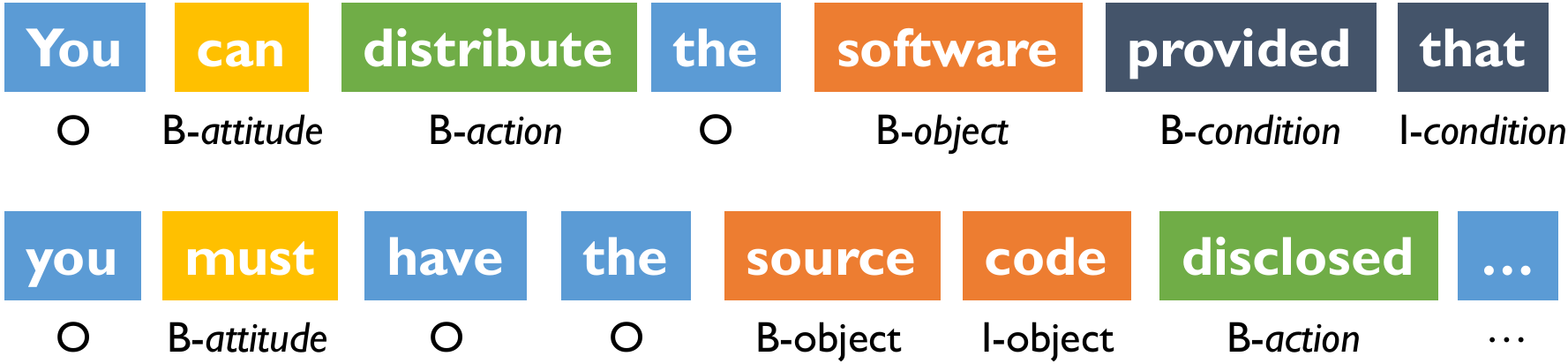}
    %\vspace{-2mm}
    \caption{Illustration of the BIO Mode for Entities in Licenses}
    \label{fig_entity}
\end{figure}

\revise{To extract the fine-grained entities from licenses, we employ sequence labelling to identify the beginning and last token of each entity.}
%Inspired by Named Entity Recognition (NER)~\cite{ISPRS`20-Runyu-NER} in natural language processing, we employ sequence labeling to identify and localize the aforementioned entities (i.e., actions, objects, attitudes, and conditions). 
As a fundamental step, we first split each license text into sentences and preprocess them by removing non-textual parts, checking spellings, performing stemming and morphological. \revise{In the training phase, we label each token in a license as illustrated in \Cref{fig_entity}, %In the training phase, we label each token with the BIO (Begin, Inside, and Outside) mode~\cite{CSCL`17-Reimers-BIO}. 
%As illustrated in \Cref{fig_entity}, 
where B-\textit{X} and I-\textit{X} represent the first token (i.e., the beginning token) and an inside token of an entity whose type is $X$, respectively. $O$ denotes a token outside any named entities. Note that unlike LiDetector~\cite{xu2021lidetector} which used $X$ to represent the $X^{th}$ license term, in this work, we use $X$ to represent the type of the entity. }For instance, B-\textit{object} in~\Cref{fig_entity} indicates that the token \textit{source} is at the beginning of an entity whose type is \textit{object}.

% Based on the labelled dataset, we train a probabilistic model to identify and localize named entities. Specifically, 
% we first embed words into vectors by the pre-trained Glove model~\cite{url-glove} to employ its prior knowledge on word semantics. Then, we feed the results of word embeddings into a bi-Directional
% Long Short-Term Memory (bi-LSTM) model~\cite{CSAI`19-Xia-BiLSTM}, and learn the representation of each license sentence. Finally, for each token we utilize Conditional Random Fields (CRF)~\cite{USENIX`19-Andow-Policylint} to calibrate the probability distribution over each label and make a global decision. 
{Based on the labelled dataset, we train a NER model based on spaCy~\cite{spacy} to identify and localize named entities. 
We first embed words into vector with a pre-trained \textit{roberta-base} model~\cite{Liu2019RoBERTaAR} to employ its prior knowledge on word semantics and boost the word embedding strategy using subword features. 
Then, we feed the results of word embeddings into a transition-based parser, which utilizes a deep convolutional neural network with residual connections to learn the representation of each license sentence and parse named entities.
In the inference phase, \tool sequentially predicts the label of each token in license text so as to identify and localize entities in licenses.}
For the example in~\Cref{fig_entity}, \tool extracts seven entities, i.e., two \textit{action} entities (\textit{distribute} and \textit{disclose}), two \textit{object} entities (\textit{software} and \textit{source code}), two \textit{attitude} entities (\textit{can} and \textit{must}), and a \textit{condition} entity (\textit{provide that}).

\subsubsection{{Relation Extraction}}

Relation extraction in license understanding aims to extract structured knowledge about relations between entities from unstructured license text. \revise{Unlike previous studies~\cite{xu2021lidetector,APSEC`17-Kapitsaki-termsIdentifying} which regarded a right/obligation as a whole entity, in this paper, we decompose each regulation into four types of entities, and propose a relation extraction model to reconstruct the relations between these entities for a fine-grained and structured understanding of licenses.} Since there exists no tagged dataset for entity relations in licenses and labelling is time-consuming, inspired by the promising results of prompt-tuning for few-shot tasks, we employ a prompt-based relation extraction model based on KnowPrompt~\cite{Chen2022KnowPrompt}. The basic idea is to leverage the knowledge hidden in the pre-trained language model and formalize specific tasks as cloze tasks. With the prompt template, \tool predicts the {mask} %\ling{???} 
and maps the prediction result to the classes, which are the relations between input entities. \revise{Specifically, we extract five types of relations, i.e., \textit{action-object}, \textit{action-attitude}, \textit{action-condition}, \textit{condition-action}, and \textit{others}. \textit{Action-object} represents the relation between an action and the object to be acted upon. \textit{Action-attitude} represents the relation between an action and the attitude towards this action. \textit{Action-condition} represents the relation between an action and the condition of exercising the action. \textit{Condition-action} denotes the relation between a condition and the action required in the condition. Finally, \textit{others} represent the relations other than the aforementioned relations. With these specific relations, \tool organizes the extracted entities and constructs a fine-grained understanding of licenses.}

\begin{figure}[t]
    \centering
    \includegraphics[width=0.9\linewidth]{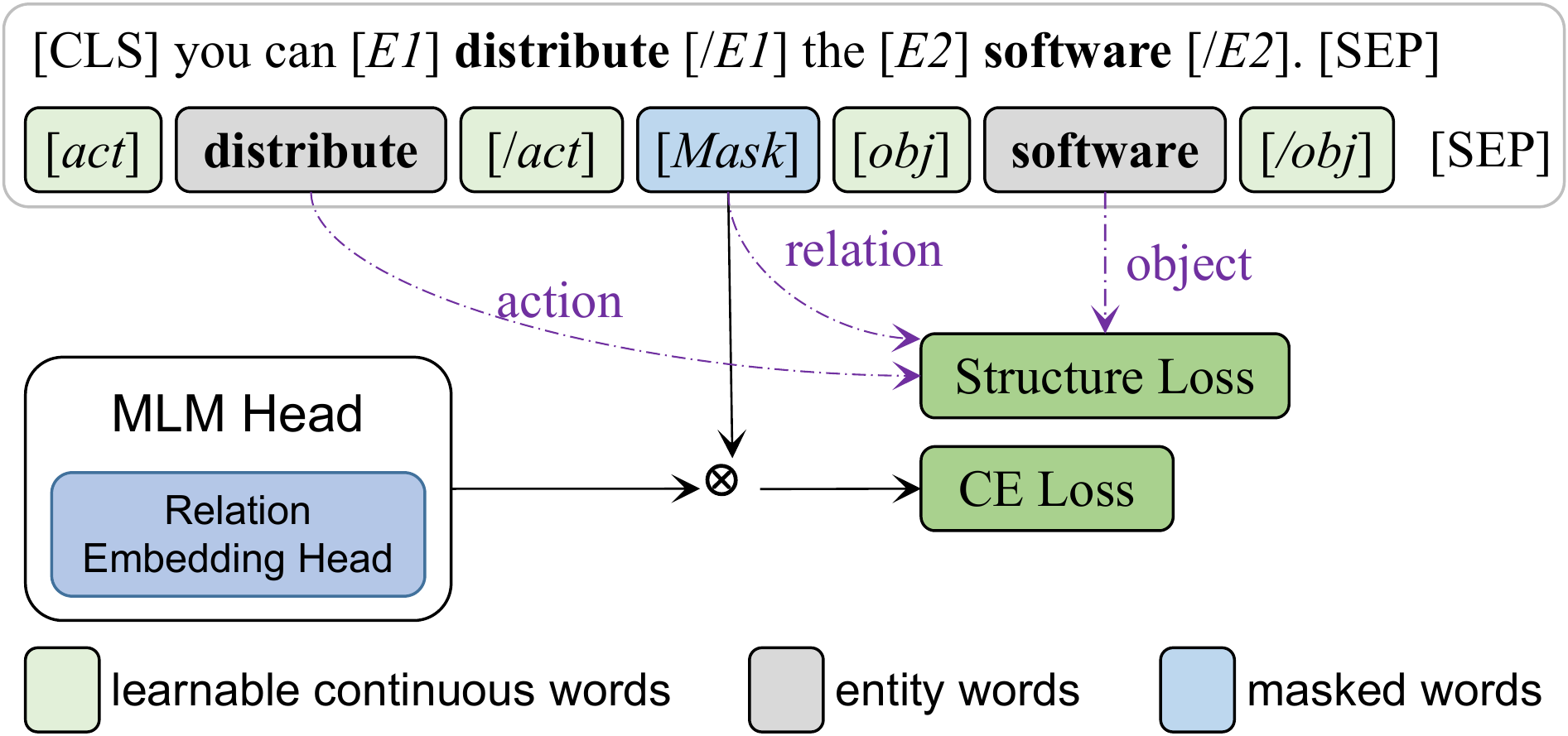}
    %\vspace{-2mm}
    \caption{Illustration of Relation Extraction from Licenses}
    \label{fig_relation}
\end{figure}

%In this paper, prompt-tuning transforms the relation extraction task into a classification task.
%Although prompt-based ideas overtake traditional ideas (e.g., fine-tuning on massive pre-trained model) on its less complex model structure and can utilize deeper language knowledge to provide quite desirable predictions, they transform almost every tasks into classification tasks due to its process of answer mapping from the answered \textit{[MASK]} to the ultimate label dictionaries. 
%Therefore, this paper applies prompt-tuning idea only on our term relation extraction task and the license term entity identification task is realized by traditional sequence tagging process of supervised training on pre-trained models because of its focus on the boundary of the term-related entities. 
% {"token": ["you", "can", "distribute", "modified", "versions", "if", "you", "keep", "track", "of", "the", "changes", "and", "the", "date", "you", "made", "them", "."], "h": {"name": "distribute", "pos": [2, 3]}, "t": {"name": "modified versions", "pos": [3, 5]}, "relation": "Action-Recipient(e1,e2)"}
\Cref{fig_relation} illustrates the model construction of relation extraction. It can be seen that given a license phrase ``\textit{you can distribute the software}'', \tool first extracts the entities \textit{can}, \textit{distribute}, and \textit{software}.
Then, taking the relations between \textit{distribute} and \textit{software} as an example, \tool constructs a prompt template as ``{[CLS] you {can} [\textit{E1}] \textbf{distribute} [\textit{/E1}] the [\textit{E2}] \textbf{software} [\textit{/E2}]. [SEP] [\textit{act}] \textbf{distribute} [/\textit{act}] [MASK] [\textit{obj}] \textbf{software} [/\textit{obj}] [SEP]}'', 
and feeds it to the Mask Language Model (MLM)~\cite{Devlin2019BERT}. 
{Here, [CLS] and [SEP] mark the input sequence and its prompt, [\textit{E1}], [\textit{/E1}], [\textit{E2}], and [\textit{/E2}] mark the beginning and end tokens of two input entities, [\textit{act}], [/\textit{act}], [\textit{obj}], and [/\textit{obj}] represent virtual types of entities learned during the prompt tuning process, and [MASK] denotes the output prediction of the MLM model. }
After that, \tool obtains the probability distribution of {[MASK]} over the vocabulary list, and maps the prediction result to the embeddings of answer words, which are the classes of relations between input entities. Finally, \tool achieves the prediction result as the relation between two entities \textit{distribute} and \textit{software}. Note that for official licenses whose rights and obligations have already been known, we only extracted fine-grained entities and relations automatically in the preparation phase, and manually checked them to construct a database that can be directly used for further analysis. The details of the database can be seen in \Cref{sec:official}.

%As for the period of license incompatibility detection, this paper mainly focus on the importance of term action and its attitude polarity (e.g., \textit{can}, \textit{cannot} and \textit{must}). Considering diverse phrases may convey the same term action meaning, several common term action types as stated in LiDetector~\cite{xu2021lidetector} are introduced and tagged in the same time during above sequence tagging process. Besides, the attitude polarity towards term action is analyzed by automatically grammar parsing and sentiment analysis as in ~\cite{xu2021lidetector}. 

\subsection{Incompatibility Issue Localization}

%In this section, we detect license incompatibility issues, and filter them by copyright holder detection, so as to localize target issues that can be resolved in the next phase.

\subsubsection{{Incompatibility Detection}}\label{sec:incompapproach}
Based on the outputs of license understanding and hierarchy extraction, %\tool detects incompatibility issues in a given project. 
{\tool identifies incompatibility issues in a given project together with the specific localization, i.e., the packages or files involved in the incompatibility issues.}
\revise{Previous works only distinguished between the project and component licenses~\cite{xu2021lidetector}, in this paper, we extract license hierarchy to obtain the licensing scope of each license, as well as the relations between multiple licenses. }%Inspired by previous studies that distinguished between the project and component licenses~\cite{xu2021lidetector}, 
\revise{Then, we can obtain \textit{parent-child} pairs from the extracted license hierarchy. Given a license $l$ and its child nodes $l_c$, we define that the license $l$ is compatible with its child nodes $l_c$ if {anyone who complies with the license $l$ will not violate any of its child nodes $l_c$}. Otherwise, there exists license incompatibility. In other words, a parent license $l$ is supposed to be equally or more restrictive than any of its child nodes $l_c$. On this basis, we propose to detect license incompatibility from bottom to up, layer by layer according to the extracted license hierarchy, so as to conduct a fine-grained license incompatibility detection. }
%we adopt a more fine-grained license incompatibility detection strategy. Specifically, based on hierarchy extraction, %we define a ``one-way compatibility'' strategy to check license compatibility, i.e., \textit{anyone who conforms to a parent license will not violate any of its child licenses}. To achieve this goal, the license at a parent node should be equally or more restrictive than all the child licenses belonging to it. 
%\revise{we examine the license heirarchy from bottom to }

%Based on the aforementioned definition, we detect license incompatibility issues from bottom to top according to the license hierarchy of the project. 
For each \textit{parent-child} pair of licenses, we extract the fine-grained entities and relations as mentioned in \Cref{sec:und}. \revise{Then, a fundamental step for incompatibility analysis is semantic alignment. Specifically, for each \textit{attidue} entity (e.g., ``are not allowed''), we follow previous works~\cite{xu2021lidetector,tldrlegal} to assign it a label (i.e., CAN, CANNOT, or MUST) for further analysis.} We also classify the entities of \textit{actions} into 23 categories as previous studies~\cite{APSEC`17-Kapitsaki-termsIdentifying,kapitsaki2019modeling,xu2021lidetector}. Each group represents a type of actions that licensees may do. To this end, we trained a classification model with the Support Vector Machine algorithm~\cite{hearst1998support} based on the dataset provided in~\cite{xu2021lidetector}. To conduct an accurate analysis, we further distinguish the same \textit{actions} with different \textit{objects} into different groups. For example, \textit{distribute source code} and \textit{distribute binaries} are regarded as two different regulations. 

By this means, we group the regulations from the \textit{parent-child} pair of licenses. Each group contains the regulations about the same \textit{action-object} but multiple attitudes, and \tool detects whether there exist conflict attitudes between a parent license and its child license towards the same \textit{action-object}.
%After aligning license terms represented by \textit{action-object} pairs, we detect whether there exist conflict attitudes between a parent license and its child license towards the same \textit{action-object}. %Given the ``one-way-compatibility'' definition, a parent license is compatible with a child license only if it is equally or more restrictive than the child license. 
\revise{For example, if \tool infers from a parent license that states \textit{can distribute source code}, while infers from one of its child licenses that claims \textit{cannot distribute source code}, in this case, anyone who conforms to the parent license may violate at least one of its child licenses by distributing source code.}
%For example, towards the same \textit{action-object}, the \textit{attitude} MUST is stricter than CAN, hence anyone who conforms to MUST will not violate CAN. 
%Moreover, we observe that there might be some license terms that only appear in one license (e.g., the parent license). Following previous studies~\cite{xu2021lidetector}, we take the absence of license terms as rights reserved, and categorize \textit{actions} into two types, i.e., right and obligation. For the right-related \textit{actions}, their absence in a license indicates that licensees CANNOT exercise these rights. For the obligations, the absence of them are marked with CAN by default. 
As for the \textit{conditions}, if a \textit{condition} entity is extracted as well as its relations with other entities, we consider the conditional cases separately, i.e., separately assume two cases where the condition is $True$ or $False$, respectively. %Specifically, we first assume the condition is $True$ and update the attitudes towards the \textit{actions} in the main and subordinate clauses for incompatibility detection. Then, we assume the condition is $False$, update the attitudes, and check incompatibility in this case.
{By this means, we can identify incompatibility issues and the detailed information about such issues, including the incompatible parent and child nodes (i.e., licenses and the packages/files where they are located), the incompatible \textit{action-object} pairs, the \textit{attitudes} and \textit{conditions} (optional) of the involved licenses, which are further utilized for license incompatibility resolution.} Note that as previous studies~\cite{url-choosealicense,xu2021lidetector}, the absence of a project license implies all rights are reserved, i.e., nobody can copy, distribute, or modify the work. In this case, \tool detects whether the licenses at the first level of the hierarchy can be incorporated into the same project without license violations, and reports incompatible licenses.

\subsubsection{{Copyright Holder Filtering}}
%\ling{why here. may move to 2.4?}
% As the research of Vendome .et al~\cite{Vendome2016Assisting}, the copyright holders 
\revise{Previous research mainly focused on the project license whose licensing scope is the whole project, while ignoring other licenses in the project~\cite{liu2019predicting,kapitsaki2019modeling,xu2021lidetector}. 
 However, in this paper, we notice that there are often more than one license that can be changed for incompatibility resolution.} The intuition behind is that if a license has the same copyright holder with the project license, the project owner ought to have the right to modify or replace it. 
Actually, it is a common practice for developers of large-scale OSS to place their own license claimers in the modules implemented by themselves for copyright protection and ownership assertion of these modules. 

For the above reason, we propose to identify copyright holders so as to enlarge the scope of licenses that can be modified for license incompatibility resolution. Since we assume that a project owner can always modify the project license, \tool first identifies the copyright holder stated by the project license. Then, it examines all licenses involved in license incompatibility issues, so as to localize the module licenses (also known as component licenses) with the same copyright holder as the project license, which indicates that these licenses can also be modified by the project owner. 

Specifically, we first identify the license sentences that claim copyright holders. Since the expressions of copyright holders are relatively fixed, we define a set of phrases as the signals of copyright claim, for instance, \textit{copyright (c)} and \textit{authored by}. The specific expressions can be seen online~\cite{ourwebsite}. We detect these keywords by regular matching. 
After identifying the copyright-related sentences, we utilize Stanford CoreNLP~\cite{url-corenlp} to recognize named entities related to \textit{person names} and \textit{organization names} as the copyright holder of the input license.

For the example in \Cref{withHie}, \tool first identified the copyright holder \textit{Chris Tabor} claimed in the project license. Then, it detected 10 component licenses inside the project that have the same copyright holder with the project license, such as the inline licenses (i.e., the MIT license) at the beginning of source code files \textit{db.py} and \textit{wordcloud.py}. In this case, there exist more than one license that can be modified to resolve license incompatibility issues.

\subsection{License Incompatibility Resolution}

\SetKwInput{KwInput}{Input}
\SetKwInput{KwInput}{Input}
\SetKw{Let}{let}
\SetKw{Continue}{continue}
\SetKw{Break}{break}
\SetKw{Create}{CREATE}
\begin{algorithm2e}[t]
%\color{blue}
    \footnotesize
	\setcounter{AlgoLine}{0}
	\caption{{License Incompatibility Resolution}}
	\label{algo}
	\DontPrintSemicolon
	\SetCommentSty{mycommfont}
	%\SetAlgoLined
	{
	   % \KwIn{$l_1<t_1, atti_{t1}>$ and $l_2<t_2, atti_{t2}>$: A pair of licenses with extracted terms and attitudes}
	   % \KwIn{$l$, $l_{prt}$, $LC \gets l_{cdn}$: A license to repair and its parent license and numerous child licenses, each of which is composed of 
	   %$<t, atti_{t}, cond_{t}, >$}
	    %\KwIn{$LR \gets l_{repair}$: the set of licenses with incompatibility issues repair in given OSS project after $CompDetect()$, each of which is composed of $<t, atti_{t}, cond_{t}, >$ based on results of $LiUnderStand()$ }
	    \KwIn{$X$: the target project with license incompatibility issues}
    	\KwOut{$S$: the resolution suggestions of the incompatibility issues}
    	
    	\SetKwProg{Fn}{Function}{:}{}
                    $LR \gets detectCompt(X)$ \tcp*[h]{obtain incompatible licenses that can be changed.}\;
                    $S \gets \emptyset$ \tcp*[h]{resolution of the project.}\;
    	        \ForEach{$l \in LR$}{
    	            \tcp*[h]{iterate licenses in $LR$ {from bottom to top}.} \;
    	            $l_{p} \gets \text{getParent}(X, l)$  \;
                        $r_p \gets \text{getReq}(l_p)$ \;
                        $l_c \gets \text{getChildren}(X, l)$ \;
                        $r_c \gets \emptyset$ \;
                        $f_l \gets False$ \tcp*[h]{initialize the flag for exception.} \;
                        \ForEach{$c \in l_c$}{
                         $r \gets \text{getReq}(c)$ \;
                         $r_c, f \gets \text{mergeReq}(r_c, r)$ \\\tcp*[h]{merge the requirements of child nodes.}\;
                         $f_l \gets f_l \lor f$ \tcp*[h]{if there are conflicts between child nodes.}\; 
    	        }
    	            $R_l \gets \text{resolve}(r_p, r_c) $ \tcp*[h]{resolve the constraints for license $l$, i.e., no more restrictive than $r_p$ but no more loose than $r_c$.} \;
                        \eIf{$R_l$ is None}{
                        \Continue \tcp*[h]{if the constraints cannot be resolved, continue.} \;
                    }
                 {
                        $s_l.o \gets \text{checkOfficial}(R_l)$ \;
                        \If{$s_l.o$ is None}{
                            $s_l.c \gets \text{generateCustom}(R_l)$ \;
                            \If{$f_l$ is TRUE}{
                                $s_l.c \gets \text{addException}(R_l, s_l.c)$
                        }
                }   
                $S \gets S \cup s_l$
    	        }
             \Return $S$
    	        }
    }

\end{algorithm2e}

\iffalse
   	        \ForEach{$l_i \in LR$}{
    	            \tcp*[h]{choose specific repair solution for each license to repair. } \;
    	            
    	            $isRecomd$, $recomdLi$ = checkOfficialLi($Need_i$, $L_{offic}$) \;
    	            \tcp*[h]{iterate all official licenses based on their interpreted information, and check 
    	            whether the repair constraint of this license can be satisfied. } \;
    	            \eIf{$isRecomd$ is $False$}{
    	                
    	                \tcp*[h]{generate a custom license based on the term essentials interpreted. } \;
    	                $customLi$ = geneCustomLi($Need_i$) \;
    	                
    	                $isExcep$, $excepTerms$ = checkExceptionNeeds($Need_i$) \;
    	                \eIf{$isExcep$ is $True$}{
    	                    \tcp*[h]{there is some exception constraints and we must add exceptions to the base custom licenses.} \;
    	                    \ForEach{$t_{excep} \in excepTerms$}{
    	                        \ForEach{$atti \in t_{excep}.Atti$}{
    	                            $filepath_{atti}$ = findAttiSource($atti$, $t_{excep}$, $l_i$, $OSS$) \;
    	                            $exceptLi$ = updateExceptionLi($customLi$, $filepath_{atti}$) \;
    	                        }
    	                    }
    	                    \Return $exceptLi$ \;
    	                }{
    	                    \Return $customLi$ \;
    	                }
    	                
    	            }{
    	                \Return $recomdLi$ \;
    	            }
    	       }
\fi

\subsubsection{{License Constraint Resolution}}\label{sec:constraint}

As described in \Cref{algo}, 
%after license incompatibility %detection \ling{identification?} and copyright holder identification, 
{taking the project with license incompatibility issues as input,}
\tool obtains a set of problematic licenses (denoted by $LR$) which can be modified or replaced to handle the incompatibility issues (Line 1). Based on hierarchy extraction mentioned in \Cref{sec:hie}, \tool iterates the licenses in $LR$ from bottom to 
top for license incompatibility resolution (\revise{Lines 3--23}). Specifically, for each license in $LR$, \tool first obtains its parent and child nodes, and then computes the requirements brought by them (\revise{Lines 4--12}). Since there may exists multiple child nodes for a license, we merge the requirements from all child nodes by extracting the most restrictive attitude towards each \textit{action} (Line 11). Note that to obtain constraints, we only take into account the licenses without the same copyright holder with the project license (i.e., the licenses that cannot be modified by the software owner). The intuition behind is that \textit{licenses with the same copyright holder of the project license can be changed to address incompatibility issues, and thus should not be the constraints for the current license/node}. For the example shown in \Cref{withHie}, there exist three licenses that can be modified (i.e., one project license and two inline licenses stated in \textit{db.py} and \textit{wordcloud.py}). In this case, \tool resolves the incompatibility issues from bottom to top, and starts from a MIT license in a Python file. Then, it computes the requirements from two child licenses, which are imported by the third-party software packages. As for the parent license, it can be seen that although the current license should be no more restrictive than its parent node (i.e., the project license in this case), however, the project license can be changed later to satisfy the requirements of compatibility, which should not be considered as a source of constraints for the current node.

Based on the license \textit{actions}, \textit{attitudes}, \textit{conditions}, and \textit{objects} extracted in the phase of license understanding, \tool resolves the constraints brought by its parent and child nodes. The basic idea is that the current license $l$ should be no more restrictive than its parent license $l_p$ but no more loose than its child nodes $l_c$. 
\revise{Specifically, given a license $l$, \tool merges the requirements from its child nodes by grouping the \textit{action-object} pairs mentioned in \Cref{sec:incompapproach}, so that the requirements towards the same \textit{action-object} are within the same group. Then, for each \textit{action-object} pair, \tool merges the requirements from the children by obtaining the most restrictive attitude towards the same \textit{action-object}, which serves as a constraint for this \textit{action-object}.} Similarly, \tool obtains the requirements from the parent license, which serve as another boundary of the constraints. By this means, \tool achieves the constraints from the context and stores them in $R_l$.

\subsubsection{{Official License Recommendation}} \label{sec:official}

After license constraint resolution, we obtain the constraints for a current node, which is a license involved in license incompatibility and can be changed by the project owner. However, we note that not all constraints from the context can be resolved. According to the ``one-way-compatibility'' definition, given a license, if it has a child license which is more permissive than its parent license, the requirements from the context of the current license (i.e., its parent and child licenses) cannot be satisfied simultaneously. In this case, the constraints for fixing the incompatibility issue cannot be resolved, \tool moves on to the next target license. Hence, as described in \Cref{algo}, \tool only provides resolutions for licenses whose constraints from their context can be resolved by \tool (\revise{Lines 17--22}). Specifically, after resolving the constraints, \tool first checks whether there exist official licenses that can satisfy the requirements. In this paper, official licenses represent the licenses from the Software Package Data Exchange (SPDX)~\cite{spdx} which are publicly available. Note that the high priority of recommending official licenses comes from an empirical study that reported official licenses counted for more than half of the investigated licenses rather than custom licenses~\cite{xu2021lidetector}. However, \tool also provides settings for users to set priority of resolutions.

To recommend official licenses for license incompatibility resolution, we constructed a database of official licenses whose rights and obligations have already been known by previous studies~\cite{tldrlegal,xu2021lidetector}. We manually checked the license terms labelled by previous studies (e.g., CAN \textit{distribute}), extracted the fine-grained entities and relations mentioned in \Cref{sec:und} using \tool, and verified the labels by three authors and a lawyer to construct the database. Finally, when more than one official licenses satisfy the constraints, we rank these licenses according to their similarities to the original license via cosine similarity. %\sihan{Note that for each incompatible license to be resolved, we only take into consideration its parent and child licenses , without  to other licenses such as the grand-parent licenses.}

\subsubsection{{Custom License Generation}}\label{sec:custom}

Except official licenses, developers are also allowed to customize their own licenses, which are named by \textit{custom licenses} in this paper. As described in \Cref{algo}, given a set of constraints, it is possible that none of existing official licenses can satisfy the constraints $R_l$. In this case, as an alternative way, we propose to generate a custom license for the project owner to choose (\revise{Lines 18--21}). We note that although there are no conventions for custom licenses to comply with, however, as a type of contracts between the owner and users of projects, the expressions of licenses need to be precise and strict in most cases. For this reason, despite the remarkable progress of text generation techniques in natural language processing~\cite{yu2022survey}, we do not utilize machine learning-based methods for license text generation. Instead, in this paper, we propose a template-based approach to generate custom licenses according to the constraints and fine-grained information obtained in the aforementioned steps. 

As displayed in \Cref{algo}, $R_l$ represents the results of constraint resolution for a target license $l$. Then, \tool \textit{automatically} generates a custom license by organizing the fine-grained entities and their relations stored in $R_l$. Specifically, for each \textit{action} in $R_l$, if there exists an \textit{object} and an \textit{action-object} relation between them, then \tool concatenates them directly or with a preposition determined by {CoreNLP}~\cite{url-corenlp}. As for the \textit{attitude}, we note that the results of constraint resolution may contain multiple attitudes toward the same \textit{action-object}. For instance, given a parent node with the {attitude} MUST and a child node with the {attitude} CAN towards the same \textit{action-object}, the result of constraint resolution for the current node contains a set of \textit{attitudes} [CAN, MUST] for this \textit{action-object} pair, which is no more restrictive than the parent node and no more permissive than the child node. In this case, \tool uses the most restrictive \textit{attitude} in the set by default. In addition, when a \textit{condition} entity is found along with its relation with the \textit{action} (i.e., \textit{action-condition}), \tool adds the \textit{condition} at the end of the current sentence, equipped with the \textit{action}, \textit{attitude}, and \textit{object} (optional) orderly found by the relations \textit{condition-action}, \textit{action-attitude}, and \textit{action-object}. \revise{For instance, after extracting an \textit{action} entity ``modify'' and an \textit{object} entity ``source code'', as well as the \textit{action-object} relation between them, \tool concatenates them and generates ``{modify source code}'' as a result. If there exists a \textit{condition} entity ``as long as'' in $R_l$ and an \textit{action-condition} relation between ``{modify}'' and ``as long as'', \tool extends the statement and generates ``{modify source code as long as ...}''. Finally, to generate a clear and concise license, \tool merges the sentences with the same \textit{objects} (e.g., you can \textit{distribute} and \textit{modify} copies of software)}.

\smallskip
\noindent \textbf{Exceptions}. In this paper, we observe that there may exist some conflicts between the requirements of multiple child licenses which cannot be modified by the project owner. A natural solution to this problem is to replace the third-party packages involved in the conflicts. However, since the replacement of a third-party package requires substantial efforts due to the complex package dependency, we also provide an alternative way by attaching an \textit{exception}, which grants an exception to the license or some additional permissions (e.g., LLVM Exceptions~\cite{url-llvm}). Specifically, \tool first localizes the files, packages or directories of the conflict licenses. Then, towards the conflict \textit{action} and \textit{object} (optional), it assigns different attitudes with different licensing scopes in the same license. By this means, \tool attaches an \textit{exception} to the target license, so as to mitigate the incompatibility issue.

\section{Evaluation} \label{sec:evaluation}

% the number of copyright holder

In this section, we present the evaluation results of \tool on real-world datasets and answer the following research questions. % Specifically, we present the experimental results of \tool in three phases, i.e., license understanding, incompatibility issue localization, and license incompatibility resolution, and answer the following research 

    \textbf{RQ1}: Can \tool provide a fine-grained understanding of license texts by effectively extracting entities and their relations?
    
    \textbf{RQ2}: Can \tool outperform state-of-the-art approaches in detecting and localizing license incompatibility issues?
    
    \textbf{RQ3}: How effective is \tool for resolving license incompatibility issues in real-world OSS?

\subsection{Evaluation on License Understanding}
In this section, we evaluate the performance of \tool on fine-grained entity extraction and relation extraction.
%\subsubsection{\textbf{Setup}}  
For both tasks we performed labelling, training, and testing in \textbf{sentences}. The sentence datasets were constructed based on the dataset from a previous study~\cite{xu2021lidetector}. \revise{To reduce bias, we did not set aside any licenses when evaluating the effectiveness of license understanding provided by \tool. In total, there were 212 official licenses and 188 custom licenses, comprising 21,844 license sentences  (i.e., {11,973} from \textit{tldrlegal}~\cite{tldrlegal} and {9,871} from GitHub). We manually labelled the entities and relations. During labeling, we took Fleiss' Kappa to assess the reliability of agreement among the raters. The result is 0.83, indicating that the raters reach high agreement.} We randomly split each dataset into the training, validation and testing datasets by 3:1:1. 

\subsubsection{{Entity Extraction}.}
% To evaluate the performance of \tool for entity extraction, we compare it with two state-of-the-art tools (i.e., FOSS-LTE~\cite{APSEC`17-Kapitsaki-termsIdentifying} and LiDetector~\cite{xu2021lidetector}) and two natural language processing techniques (i.e., regular matching~\cite{USENIX`19-Andow-Policylint} and semantic similarity~\cite{goldberg2014word2vec}) as follows:

% \begin{itemize}[leftmargin=*]
%     \item \textbf{FOSS-LTE}~\cite{APSEC`17-Kapitsaki-termsIdentifying}, which employed a topic model named Latent Dirichlet Allocation (LDA)~\cite{MTA`19-Jelodar-LDA} to identify license terms.
%     \item \textbf{LiDetector}~\cite{xu2021lidetector}, which utilized sequence labelling and sentiment analysis to identify license term entities (similar to the \textit{actions} in this paper) and \textit{attitudes}, respectively.
%     \item \textbf{Regular matching}~\cite{USENIX`19-Andow-Policylint}, which was implemented by predefining a set of keyword patterns for entity extraction. 
%     \item \textbf{Semantic similarity}~\cite{goldberg2014word2vec}, which utilized the word2vec pre-trained language model~\cite{mikolov2013distributed,mikolov2013efficient} to measure the cosine similarity between the tokens and predefined keywords. 
% \end{itemize}

To evaluate the performance of \tool for entity extraction, we compare it with two state-of-the-art tools (i.e., \textbf{FOSS-LTE}~\cite{APSEC`17-Kapitsaki-termsIdentifying} and \textbf{LiDetector}~\cite{xu2021lidetector}) and two natural language processing techniques (i.e., \textbf{regular matching}~\cite{USENIX`19-Andow-Policylint} and \textbf{semantic similarity}~\cite{goldberg2014word2vec}).
\revise{Specifically, {FOSS-LTE}~\cite{APSEC`17-Kapitsaki-termsIdentifying} employed a topic model named Latent Dirichlet Allocation (LDA)~\cite{MTA`19-Jelodar-LDA} to identify license terms.
{LiDetector}~\cite{xu2021lidetector} utilized sequence labelling and sentiment analysis to identify license term entities (similar to the \textit{actions} in this paper) and {attitudes}.
{Regular matching}~\cite{USENIX`19-Andow-Policylint} was implemented by pre-defining a set of keyword patterns for entity extraction.
{Semantic similarity}~\cite{goldberg2014word2vec} utilized the word2vec pre-trained language model~\cite{mikolov2013distributed,mikolov2013efficient} to measure the cosine similarity between the tokens and predefined keywords.}
Note that the outputs of FOSS-LTE are license term phrases mapped from topic sentences (e.g., MayGrantPatents). To conduct a fair comparison of license understanding, we split the outputs of FOSS-LTE into four types of entities as other baselines. Similarly, LiDetector naturally extracts license terms (e.g., Commercial Use), attitudes (e.g., CANNOT), and conditions (e.g., \textit{provided that}) from license texts, which can be compared with the \textit{actions}, \textit{attitudes}, and \textit{conditions} in this paper. As for regular matching and semantic similarity, we analyzed 900 license sentences and summarized a set of keyword patterns for each type of entities as shown online~\cite{ourwebsite}. 

%\subsubsection{\textbf{Results}}

\begin{table}[]
\centering
\small
\caption{Comparison on Fine-grained Entity Extraction}
\scalebox{0.73}{\begin{tabular}{ccccccccccc}
\toprule
\multirow{2}{*}{\textbf{Tool}} & \multicolumn{2}{c}{\textbf{Actions}}              & \multicolumn{2}{c}{\textbf{Objects}}           & \multicolumn{2}{c}{\textbf{Attitudes}}             & \multicolumn{2}{c}{\textbf{Conditions}}    & \multicolumn{2}{c}{{\revise{\textbf{Total}}}}       \\\cline{2-11}
                                 & \textbf{P (\%)} & \textbf{R (\%)} & \textbf{P (\%)} & \textbf{R (\%)} &  \textbf{P (\%)} & \textbf{R (\%)} &  \textbf{P (\%)} & \textbf{R (\%)} & \revise{\textbf{P (\%)}} & \revise{\textbf{R (\%)}} \\
\midrule
{Reg. Match.} & 54.76             & 65.07             & 5.32            & 22.38             & 2.58             & 47.61             & 46.87             & 61.64      & \revise{14.74} & \revise{49.41}        \\
Similarity              & 50.81             & 59.96            & 3.18             & 12.27             & 2.35             & 40.47             & 46.86             & 61.65      & \revise{14.28} & \revise{42.33}       \\
FOSS-LTE                         & 35.98             & 23.12             & 5.76             & 19.13             & 3.06             & 26.19             & 66.67            & 21.92       & \revise{14.30} & \revise{21.12}      \\
LiDetector                       & 43.14             & 55.79             & -             & -            & 74.74             & 56.35             & {90.25}             & 86.30         & \revise{47.79} & \revise{32.80}     \\
\textbf{\tool}                             & \textbf{80.79}             & \textbf{84.86}             & \textbf{59.23}             & \textbf{63.83}             & \textbf{78.69}             & \textbf{76.80}             & \textbf{95.45}             & \textbf{87.50}         & \textbf{\revise{74.15}} & \textbf{\revise{77.31}}     \\
\bottomrule
\label{tab:entity}
\end{tabular}}
\end{table}

\Cref{tab:entity} displays the experimental results of entity extraction for five approaches. Note that since LiDetector does not extract the \textit{objects} of \textit{actions} in license texts, the results of extracting \textit{objects} by LiDetector are denoted by ``-''. In total, \tool achieves 74.07\% precision and 77.36\% recall for entity extraction. It can be seen that the precision of \tool is much higher than that of the baselines by at least 26.03\%, 53.47\%, 3.95\%, and 5.2\% when extracting \textit{actions}, \textit{objects}, \textit{attitudes}, and \textit{conditions}, respectively. It indicates that \tool can precisely extract entities in license texts without much noise. However, regular matching and similarity-based approaches achieve low precision especially for \textit{objects} and \textit{attitudes}. By analyzing the results, a possible explanation could be the ambiguous expressions for the two types of entities. 
Moreover, the recall scores achieved by \tool are also higher than those of four baselines by at least 19.79\%, 41.45\%, 29.19\%, and 1.2\% when extracting \textit{actions}, \textit{objects}, \textit{attitudes}, and \textit{conditions}, respectively. %\tool achieved the highest performance on \textit{conditions}. A possible reason is that the expressions of \textit{conditions} are less flexible than other types of entities. 
Among four baselines, LiDetector achieves a competitive performance with \tool on extracting \textit{attitudes} and \textit{conditions}. 
However, the precision and recall scores of LiDetector on extracting \textit{actions} are lower than those of \tool. By analyzing the entities extracted by LiDetector, we found that {the entities extracted by LiDetector are more coarse-grained, sometimes along with its \textit{objects} or \textit{attitudes}, which made them longer than the entities extracted by other approaches, which decreased the evaluation results of LiDetector.}

\subsubsection{{Relation Extraction}}
\begin{figure}[]
    \centering
    \includegraphics[width=0.98\linewidth]{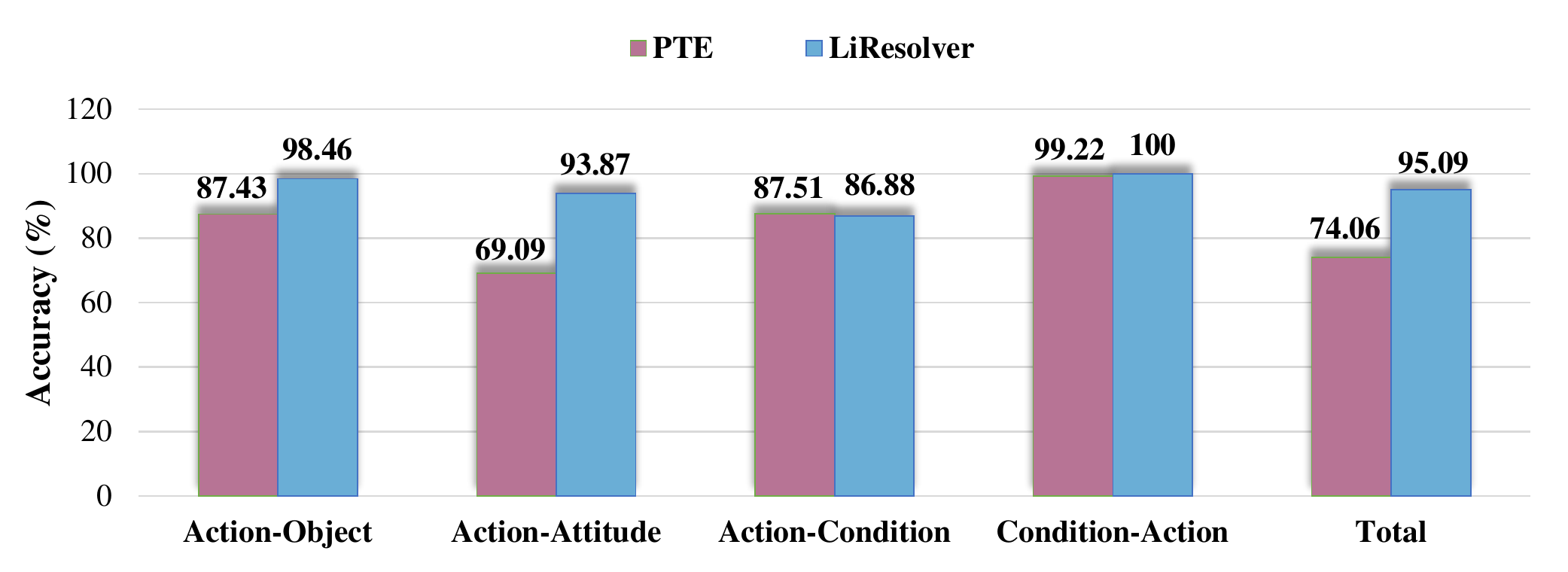}
    %\vspace{-2mm}
    \caption{Accuracy of Relation Extraction}
    \label{fig:relation}
\end{figure}

\Cref{fig:relation} shows the performance of two approaches on relation extraction. Since FOSS-LTE~\cite{APSEC`17-Kapitsaki-termsIdentifying} and LiDetector~\cite{xu2021lidetector} do not extract relations between entities, we did not compare \tool with them. Instead, we implemented a heuristic approach based on the positions and types of entities (denoted by \textbf{PTE}). Specifically, it determined entity relations according to their types and relative positions in the same sentence. For instance, \textit{action-object} indicates an \textit{action} in front of \textit{object}. Similarly, \textit{action-condition} and \textit{condition-action} were also distinguished by the relative positions of the \textit{action} and \textit{condition}. %As for grammar parsing, it first utilized CoreNLP~\cite{url-corenlp} for dependency parsing, which provided semantic dependencies between phrases in a sentence. Then, it classified the relations between entities according to these dependencies. For example, the relation between a pair of entities with the \textit{obj} dependency is considered as \textit{action-object}. The detailed mappings between semantic dependencies and entity relations are available online~\cite{ourwebsite}.
Finally, to conduct a fair comparison, we fed both approaches the same pairs of entities as input. 
%
%In \Cref{tab:realtion}, we use \textit{Act}, \textit{Obj}, \textit{Att}, and \textit{Con} to represent \textit{action}, \textit{object}, \textit{attitude}, and \textit{condition}, respectively. 
Since the input of entity relation is a pair of entities, there exist no false negatives and thus we evaluate the performance of two approaches only with the accuracy metric. It can be seen that \tool can effectively identify entity relations, achieving 95.69\% accuracy in total. We can also observe that the performance of \tool is superior to that of the heuristic baseline for all types of relations. The baseline only achieves comparable performance with \tool on \textit{action-condition} and \textit{condition-action}. It indicates that the relation between an \textit{action} and a \textit{condition} can be easier inferred by their relative positions in the sentence compared to other types of entity relations. However, since PTE requires the types of entities as input, its performance relies on the accuracy of entity extraction, while \tool has no requirements for such information.

%In detail, a pair of entities with \textit{obj} dependency is considered as \textit{action-recipient}, a pair of entities with \textit{aux, advmod} dependencies is considered as \textit{action-attitude}, and a pair of entities with \textit{acl, advcl} dependencies (usually from \textit{action} to \textit{condition}) is considered as existing \textit{condition} relation, the dependency direction of which decides it falls into \textit{action-condition} or \textit{condition-action}. 

\noindent \fbox{
\parbox{0.95\linewidth}{
\textbf{Answer to RQ1:} 
\tool can effectively extract fine-grained entities and their relations from license texts, with 74.07\% precision and 77.36\% recall for entity extraction, 95.09\% accuracy for relation extraction, which provides detailed information for license incompatibility localization and resolution.
}}

\iffalse
\begin{table}
	\centering
	\small
	\caption{Accuracy of Relation Extraction \sihan{to update}}
	%\scalebox{0.9}{
    \begin{tabular}{ccccc|c}
		\toprule
		%\hline
		\textbf{Method} & \textbf{Act-Obj} & \textbf{Act-Att} & \textbf{Act-Con} & \textbf{Con-Act} & \textbf{Total}\\
		\hline
		 PTE  & 87.43  & 69.09 & 87.51 & 99.22 & 74.06\\
		%Grammar Parsing  & 42.77  & 29.29 & 1.01 & 0.00 & 16.80\\
		\textbf{\tool}  & \textbf{98.46}  & \textbf{93.87} & \textbf{86.88} & \textbf{100.00} & \textbf{95.69}\\
		%\hline
		\bottomrule
		\label{tab:realtion}
	\end{tabular}%}
\end{table}
\fi

\iffalse
\begin{table}
	\centering
	\small
	\caption{Comparison on License Understanding \sihan{to update}}
	\scalebox{0.9}{\begin{tabular}{ccccc}
		\toprule
		%\hline
		\textbf{Entity Extraction} & \textbf{Relation Extraction}& P & R & C\\
		\midrule
		Regular Matching  & Regular Matching  &XX  & XX & XX \\
		Semantic Similarity  & Regular Matching  & XX  & XX & XX \\
		FOSS-LTE  & FOSS-LTE  & XX  & XX & XX \\
		LiDetector  & LiDetector  & XX  & XX & XX \\
		\textbf{\tool}  &\textbf{\tool}  & \textbf{XX}  & \textbf{XX} & \textbf{XX} \\
		%\hline
		\bottomrule
		\label{tab:realtion}
	\end{tabular}}
\end{table}
\fi

\subsection{Evaluation on Incompatibility Localization}\label{sec:incomp}
%All the techniques and tools were evaluated on the same testing dataset. 
%In this section, we evaluate the ability of \tool on detecting and localizing license incompatibility issues.
\subsubsection{{Setup}}
We evaluate the performance of \tool on 300 real-world projects on GitHub. Specifically, we first crawled 10,000 popular projects with more than 1,000 stars. Then, we randomly selected 300 projects and constructed a ground-truth dataset. Three authors and a lawyer were involved in labelling and verifying the dataset which has been made publicly available online~\cite{ourwebsite}. In total, we extracted {81,402} licenses for incompatibility analysis. %, including \sihan{XXX} unique official licenses and \sihan{XXX} custom licenses.  
We compare \tool with four state-of-the-art tools, i.e., the SPDX Violation Tools (\textbf{SPDX-VT})~\cite{jss`17-Kapitsaki-SPDX}, \textbf{Ninka}~\cite{ase`10-German-Ninka} equipped with \textit{tldrlegal}~\cite{tldrlegal}, {\textbf{Librariesio}~\cite{url-librariesio}, and \textbf{LiDetector}~\cite{xu2021lidetector}}. {SPDX-VT}~\cite{jss`17-Kapitsaki-SPDX}, which designed a directed graph to represent the compatibility relations between a set of licenses in Software Package Data Exchange (SPDX)~\cite{spdx}. {Ninka}~\cite{ase`10-German-Ninka}, which identified licenses with a sentence-matching method. Combined with the summarized rights and obligations provided by \textit{tldrlegal}~\cite{tldrlegal}, Ninka can be utilized to detect incompatibilities between licenses. {Librariesio}~\cite{url-librariesio}, which checked compatibility between a predefined set of SPDX licenses. {LiDetector}~\cite{xu2021lidetector}, which firstly introduced a learning-based method to automatically identify license terms from arbitrary licenses and detect license incompatibilities. 
\iffalse
\begin{itemize}[leftmargin=*]
    \item \textbf{SPDX-VT}~\cite{jss`17-Kapitsaki-SPDX}, which designed a directed graph to represent the compatibility relations between a set of licenses in Software Package Data Exchange (SPDX)~\cite{spdx}.
    \item \textbf{Ninka}~\cite{ase`10-German-Ninka}, which identified licenses with a sentence-matching method. Combined with the summarized rights and obligations provided by \textit{tldrlegal}~\cite{tldrlegal}, Ninka can be utilized to detect incompatibilities between licenses.
    \item \textbf{Librariesio}~\cite{url-librariesio}, which checked compatibility between a predefined set of SPDX licenses. 
    \item \textbf{LiDetector}~\cite{xu2021lidetector}, which firstly introduced a learning-based method to automatically identify license terms from arbitrary licenses and detect license incompatibilities. 
\end{itemize}
\fi

\subsubsection{{Results}}

\Cref{tab:incom} displays the experimental results of license incompatibility localization. The results show the number of incompatible projects and incompatible issues reported by each approach. Note that in this paper, \textbf{an incompatible issue} represents a pair of licenses which are incompatible with each other and should be resolved to reduce legal risks. We also report the false positive (FP) rate and the false negative (FN) rate from both the project-level and the issue-level for five approaches. 
From~\Cref{tab:incom}, it can be seen that \tool is superior to all the baselines in terms of both FP and FN rates. Specifically, \tool reported 231 incompatible projects with 0.43\% FP rate  and no false negatives, which demonstrated its ability to detect projects with license incompatibility. 
% \gy{There are also other tools with the project-level FP to be zeros, i.e., Ninka equipped with \textit{tldrlegal} and SPDX-VT. It indicates that the projects reported by them are all incompatible projects without any false positives. However, their false negatives are also much higher than other approaches. The FN rates of Ninka and SPDX-VT are 42.86\% and 44.54\%, respectively. }

It can also be seen that Ninka equipped with \textit{tldrlegal} and SPDX-VT have low FP rates (i.e., 2.94\% and 2.63\%, respectively) but high FN rates (i.e., 42.61\% and 67.83\%).
By analyzing their detection processes and results, we found that both of them complied with strict and predefined rules for license compatibility analysis, which contributed to the low FP rates. However, relying on license identification and manually defined compatibility rules, the applications of these two approaches were limited to a small set of official licenses, while ignoring other licenses in the projects, which causes their high FN rates.
Similarly, the performance of Librariesio is close to that of Ninka and SPDX-VT.
% , with less false negatives but more false positives. 
It categorized {35} official licenses into five classes (e.g., \textit{permissive licenses}), and manually defined a heuristic rule based on the classification. It achieved a lower FN rate (i.e., 33.91\% at the project-level) than Ninka and SPDX-VT, since it took more licenses into consideration. However, due to the course-fined classification of licenses, it achieved the highest project-level FP rate among all the approaches (i.e., 11.11\%). 

Among four baselines, LiDetector achieved the closest performance to \tool for resolving license incompatibilities. Specifically, it reported 228 incompatible projects with 8 false positives and 10 false negatives, 
% By analyzing the false positive, we found that it occurred because \sihan{XXX}.
mainly caused by its coarse-grained license term extraction which did not distinguish between the same \textit{actions} with different \textit{objects} (e.g., \textit{Distribute source code} \textit{vs.} \textit{Distribute binaries}). Moreover, like other baselines, it also ignored the license hierarchy of the projects. 

\begin{table}[]
	\centering
	\footnotesize
	\caption{Comparison on License Incompatibility Localization} \label{tab:incom}
	\scalebox{0.95}{
	\begin{tabular}{cccc|ccc}
		\toprule
		%\hline
		%\textbf{Tool} & \textbf{\#Pro.} & \textbf{\#Incomp. Pro.} & \textbf{\#Incomp. Issue} & \textbf{FP} & \textbf{FN}\\
		\multirow{2}{*}{\textbf{Tool}} & \multicolumn{3}{c|}{\textbf{\#Incomp. Projects}} & \multicolumn{3}{c}{\textbf{\#Incomp. Issues}}\\\cline{2-7}
		 & {\textbf{\#Report.}} & \textbf{FP (\%)} & \textbf{FN (\%)} & \textbf{\#Report.}& \textbf{FP (\%)} & \textbf{FN (\%)}\\
		\hline
		Ninka   & 136 & 2.94 & 42.61 & 32,289 & 86.46 & 86.31 \\
		SPDX-VT   & 76 & 2.63 & 67.83 & 54,530 & 65.33 & 99.28 \\
		Librariesio   & 171 & 11.11 & 33.91 & 18,874 & 36.22 & 68.32 \\
		LiDetector   & 228 & 3.51 & 4.35 & 56,634 & 81.04 & 66.42 \\
		\textbf{\tool-Flat}   & \textbf{230} & \textbf{2.61} & \textbf{2.61} &  \textbf{25,238} & \textbf{33.81} & \textbf{47.75} \\
		\textbf{\tool}   & \textbf{231} & \textbf{0.43} & \textbf{0.00} &\textbf{33,340} & \textbf{4.09} & \textbf{0.02} \\
		%\hline
		\bottomrule
	\end{tabular}}
\end{table}

From the perspective of incompatibility issues, it can be seen from \Cref{tab:incom} that \tool reported 33,340 incompatibility issues with 4.09\% FP rate and 0.02\% FN rate, which demonstrated the ability of \tool to detect license incompatibility issues. We can also observe that the FP and FN rates of four baselines, ranging from 36.22\% to 99.28\%, are much higher than those of \tool, ranging from 0.02\% to 4.09\%. By analyzing the results, we found that the false negatives of SPDX-VT, and Librariesio were mainly due to their limited scope of licenses that can be handled. LiDetector localized more incompatibility issues than the other three baselines, due to the flexibility adaption to arbitrary licenses. However, it treated all component licenses equally, and ignored the potential hierarchy among them, which contributed to the false negatives. 
%SPDX-VT is similar as LiDetector and any pair of licenses with incompatibility relation input into its graph can lead to a incompatibility issue. 

\noindent $\bullet$ \textbf{Effects of hierarchy extraction}. 
To illustrate the usefulness of the license hierarchy on incompatibility detection, we also implemented \tool without hierarchy extraction, denoted by \tool-Flat in \Cref{tab:incom}. It can be seen that without hierarchy extraction, \tool reported 230 incompatible projects with 6 false positives and 6 false negatives,
% However, it has the same FN rate with LiDetector (i.e., 4.62\%). 
similar to the results of LiDetector.
The reason behind is that both \tool-Flat and LiDetector treated all component licenses in modules or accompanied with third-party packages equally important regardless of their licensing scopes. To be specific, \tool-Flat reported a project with licence incompatibility issues when there existed a component license that was more {restrictive} than the project license. However, \tool reported an incompatible project when there existed a child license more restrictive than its parent license. It can be inferred that given the same project and the same results of license understanding, the true positives found by \tool-Flat can theoretically be found by \tool. Nevertheless, when the parent license is not the project license, but a license in a module of the project, the parent and child licenses are regarded as two independent licenses by \tool-Flat and LiDetector, which might lead to a false negative. On the other hand, it can also be seen that the false positive rate of the issues reported by \tool-Flat is also higher than that reported by \tool. By observing the results, we found that it was due to the projects without project licenses. In this case, \tool-Flat considered all rights reserved, and created the most strict project license for incompatibility detection, which resulted in the false positives. %Moreover, it can also be seen that \tool-Flat reported fewer issues than \tool by {24.30\%}, since it ignored the incompatibility issues occurred between a parent-child pair of licenses where the parent license is not the project license.

\iffalse
\begin{table}[htp]
	\centering
	\footnotesize
	\caption{Comparison on License Incompatibility Localization} \label{tab:incom_comp}
	%\scalebox{0.9}{
	\begin{tabular}{ccccc}
		\toprule
		%\hline
		%\textbf{Tool} & \textbf{\#Pro.} & \textbf{\#Incomp. Pro.} & \textbf{\#Incomp. Issue} & \textbf{FP} & \textbf{FN}\\
		\textbf{Tool} & \makecell[c]{\textbf{\#Incomp. Pro.} \\\textbf{Reported}} & \makecell[c]{\textbf{\#Incomp.} \\ \textbf{Issues}} & \textbf{FP} & \textbf{FN}\\
		\hline
		Ninka   & 136 (136 overlap) & 32,289 (4,339 overlap) & 0 (0.00\%) & 102 (42.86\%)  \\
		SPDX-VT   & 76 (76 overlap) & - & 0 (0.00\%) & 162 (68.07\%)  \\
		Librariesio   & 171 (157 overlap) & - & 14 (8.19\%) & 81 (34.03\%)  \\
		LiDetector   & 228 (227 overlap) & 56,634 (9,835 overlap) & 1 (0.44\%) & 11 (4.62\%)  \\
		\textbf{\tool (flatten)}   & \textbf{227} & \textbf{90,556 (17,829 overlap)} & \textbf{0} (\textbf{0.00\%}) & \textbf{11} (\textbf{4.62\%})  \\
		\textbf{\tool (hierarchy)}   & \textbf{238} & \textbf{29,823} & \textbf{0} (\textbf{0.00\%}) & \textbf{0} (\textbf{0.00\%}) \\
		%\hline
		\bottomrule
	\end{tabular}%}
\end{table}
\fi

\noindent \fbox{
\parbox{0.95\linewidth}{
\textbf{Answer to RQ2:} 
\tool can effectively identify OSS with license incompatibility issues, achieving 0.43\% FP rate and no false negatives for 230 real-world incompatible OSS, outperforming four state-of-the-art approaches.
}}

\subsection{Evaluation on Incompatibility Resolution}
%In this section, we evaluate the performance of \tool on resolving license incompatibility issues. 
\subsubsection{{Setup}}
The experiment on license incompatibility resolution was conducted based on the same dataset with \Cref{sec:incomp}, which consist of {230} incompatible projects detected and verified at the previous step. For each project, there exists at least one pair of licenses which are incompatible with each other. Note that a project was regarded as resolved if all incompatibility issues could be resolved by \tool, and no new incompatibility issues were introduced into the resolved project according to \tool. In addition, to verify the correctness of resolutions provided by \tool, we also randomly selected {5\%} of the licenses recommended/generated by \tool (i.e., {1,484} licenses) for evaluation. %Note that as mentioned in \Cref{sec:custom}, we only generate a custom license when no existing official licenses can satisfy the constraints. Similarly, exceptions are attached only when the conflict licenses can not be modified by the software owner. %We conduct the evaluation which first checked whether exists an official license that can be used to resolve the current issue. Only when none of official licenses satisfied the license constraints, \tool generated a custom one as an alternative solution. Finally, {only when there were incompatible conflicts between multiple child licenses of the same parent license, \tool generated a custom license with an exception.} Note that \tool checked the copyright holders of incompatible child licenses. If a child license can be modified or changed to remove the conflict, \tool did not generate the exception.

\subsubsection{{Results}}

\begin{table}[]
    \centering\footnotesize
    \caption{License Incompatibility Resolved by \tool}
    \begin{tabular}{lrr}
    \toprule
        {\#Incompatible projects} &  230 & 100.00\%\\ 
        {\space \space-- \#Resolved projects}  &  \textbf{144} & \textbf{62.61\%}\\
        % {\space \space-- \#Unresolved incompatible projects}  &  {86} ({37.39\%})\\
        {\space \space-- \#Projects that cannot be resolved}  &  {86} & {37.39\%}\\
        % {\space \space-- \#Wrong resolutions provided by \tool}  &  {2} ({0.87\%})\\
        \midrule
        {\#Incompatible issues} &  33,340 & 100.00\%\\
        % {\#Incompatible issues found by \tool} &  \sihan{33,340} (100.00\%)\\
        {\space \space-- \#Issues resolved by recommending official licenses} &  \textbf{11,509} & \textbf{34.52\%}\\
        {\space \space-- \#Issues resolved by generating custom licenses} &  \textbf{18,185} & \textbf{54.54\%}\\
        {\space \space\space\space\space\space-- \#Custom licenses generated w/o exceptions} & 13,858  & 41.56\%\\
        {\space \space\space\space\space\space-- \#Custom licenses generated w/ exceptions} &  4,327 & 12.98\%\\
        %{$\bullet$ \#Issues resolved by custom licenses w/ exceptions} &  XXX (XXX\%)\\
        {\space \space-- \#Issues that cannot be resolved} &  3,646 & 10.94\%\\
        %{\space \space-- \#Wrong resolutions provided by \tool} &  514 (1.61\%)\\
        %{$\bullet$ \#Incompatible issues missed by \tool} &  \sihan{XXX} (XXX\%)\\
        \bottomrule
    \end{tabular}
    \label{tab:resolve}
\end{table}

\Cref{tab:resolve} shows the experimental results of \tool on license incompatibility resolution. Among the 230 incompatible projects, \tool resolved 144 projects, accounting for 62.61\% of all incompatible projects. From \Cref{tab:resolve} it can also be seen that {11,509} out of 33,340 incompatibility issues were resolved by \tool through recommending official licenses, accounting for {34.52\%} of all incompatibility issues. %Actually, there might exist more than one official license that satisfied the constraints. In this case, \tool produced a list of official licenses that can be used to replace the original license, and ranked according to their similarities to the original license via cosine similarity. 
%\sihan{Note that \tool provided resolutions to the problematic licenses one by one, since the choice of the solutions to one issue might affect the incompatibility detection and resolution for other licenses.}
%
%It can also be seen that official licenses could only resolve a small portion of license incompatibility issues. 
Since a majority of the incompatibility issues cannot be resolved via official licenses, as an alternative way, \tool also resolved issues by generating custom licenses to satisfy the constraints. In total, \tool generated {18,185} custom licenses, among which only {4,327} custom licenses were attached with specific exceptions to address with the problem when multiple child licenses of a target license are conflict with each other, and none of them can be modified by the software owner to resolve the conflict.

From \Cref{tab:resolve} it can also be observed that there are also some license incompatibility issues that cannot be resolved by \tool. The reasons are two folds. First, as mentioned in~\Cref{sec:constraint}, not all constraints from the context of a given license can be resolved; there exists some cases where the parent license of the current license is less restrictive than a child license of the current license. Second, not all licenses are allowed to be modified or replaced by the software owner. Some licenses were incorporated along with third-party software packages, and the project owner has no rights to change them. %To filter out licenses that can not be modified by the software owner, \tool detected copyright holders for all licenses involved in incompatibility issues and determined the licenses that can be modified. 
In total, {3,646} incompatible issues could not be resolved by \tool, accounting for {10.94\%} of the license incompatibility issues detected by \tool at the previous step.

Finally, to verify the correctness of resolutions provided by \tool, we randomly selected 5\% of the {29,694} licenses recommended or generated by \tool. We manually verified and cross-validated the correctness of these licenses. 
% We found that provided {510} wrong resolutions for the incompatible projects, accounting for {1.53\%} of all issues. 
By analyzing the results, {we found that {1.48\%} of verified resolutions were incorrect, mainly caused by the incorrect predictions in the phases of fine-grained entity extraction and relation extraction, which affected the license constraints for recommending official licenses and the detailed information used for generating custom licenses. }

\noindent $\bullet$ \textbf{Effects of copyright holder}. \Cref{tab:copyright} displays the results of license incompatibility resolution when allowing different ranges of licenses to be modified. In this paper, we classify licenses in a project into the project license and other licenses (also known as component licenses). Unlike previous studies that only focus on the project license, this paper observed a considerable number of component licenses which were also written by the project owner and thus could be modified for resolving license incompatibility. To investigate how the ranges of the licenses allowed to be modified affects the results of license incompatibility resolution, we implemented \tool with different settings. The experiment was conducted on the same dataset with \Cref{tab:resolve}.

From \Cref{tab:copyright} it can be seen that %when all licenses were allowed to be changed,
% \tool resolved \sihan{231} problematic projects and \sihan{33,340} incompatible issues, achieving \sihan{100\%} and \sihan{100\%} of \gy{the accuracy}, respectively. 
%\tool resolved 100\% of problematic projects and 100\% of incompatible issues. 
% By observing the results, we found that \sihan{XX\%} of the incompatibility issues were resolved by modifying the component licenses. 
%However, since many licenses are introduced accompanied with third-party software packages or modules, which could not be modified by the project owner, we also evaluated the performance of \tool when only project licenses in a given project were allowed to be changed. We can see that 
{138} projects and {26,523} incompatibility issues could be handled with only the project licenses allowed to be modified, accounting for {60.00\%} and {79.55\%} of the incompatible projects and issues, respectively. %The results indicate that more than half of the license incompatibility issues can be resolved by the project licenses. %the limitation to project licenses might contribute to many unsolved projects and issues. 
However, when only allowing the licenses with the same copyright holder with the project license to be changed, \tool only resolved {76} projects and {23,859} issues. By analyzing the results, we found that many project licenses did not explicitly state their copyright holders in their license texts, resulting in the low success rate of resolving incompatibility issues. 
In this paper, we propose to modify two types of licenses (i.e., the project licenses and the component licenses with the same copyright holder). The first intuition is that the project owner is ought to have the rights to modify or replace the project license, whose licensing scope is the whole project. The second intuition is that only a component license states the same copyright holder with as the project license states, the project owner should be allowed to modify it. From \Cref{tab:copyright} it can be seen that \tool resolved 2.61\% more projects and 9.51\% more incompatibility issues by adding copyright holder detection.
%The results also show that \tool resolved \sihan{XX} more issues by modifying component licenses with the same copyright holder with the project license.

\begin{table}[t]
	\centering
	\footnotesize
	\caption{Effects of Copyright Holder Detection} \label{tab:copyright}
	%\scalebox{0.9}{
	\begin{tabular}{ccc|cc}
		\toprule
		%\hline
		%\textbf{Tool} & \textbf{\#Pro.} & \textbf{\#Incomp. Pro.} & \textbf{\#Incomp. Issue} & \textbf{FP} & \textbf{FN}\\
		\multirow{2}{*}{\textbf{Modified}} & \multicolumn{2}{c|}{\textbf{\#Incomp. Projects}} & \multicolumn{2}{c}{\textbf{\#Incomp. Issues}}\\\cline{2-5}
		 & {\textbf{\#Resolved Pro.}} & \textbf{Suc. Rate} & \textbf{\#Resolved Issues}& \textbf{Suc. Rate} \\
		\hline
% 		\textbf{ALL}   & 231 & 100.00 & 33,340 & 100.00  \\
		{PL}   & 138 & 60.00\% & 26,523 & 79.55\%  \\
		{CH}   & 76 & 33.04\% & 23,859 & 71.56\%  \\
	  \textbf{PL+CH}  & \textbf{144} & \textbf{62.61\%} & \textbf{29,694} & \textbf{89.06\%}  \\
		%\hline
		\bottomrule
	\end{tabular}%}
%\vspace{-1.5mm}
\begin{center}
	    \footnotesize
	    %\textit{\textbf{ALL}: All licenses involved in license incompatibly issues can changed.} \\
	    {\textit{\textbf{PL}: Only the project license can be changed to resolve license incompatibility.}\\
	    \textit{\textbf{CH}: Only licenses whose copyright holder is the project owner can be changed.}}
\end{center}
\end{table}

\noindent \fbox{
\parbox{0.95\linewidth}{
\textbf{Answer to RQ3:} 
By recommending official licenses in priority and generating custom licenses as alternative solutions, \tool resolved 144 incompatible projects, among which 34.52\% were resolved by official licenses, 41.56\% and 12.98\% were resolved by custom licenses w/o and w/ exceptions, respectively.
}}

\section{Discussion} \label{sec:priority}
%\subsection{Priority settings}
%In this section, we discuss the developer feedback, lessons learned, and the limitations and threats to validity.
\subsection{Developer Feedback}
\revise{To evaluate the usefulness of \tool, we also collected the feedback from OSS developers with regard to the incompatibility detection results and suggestions. Since there exist 33,340 license incompatibility issues in the dataset described in \Cref{sec:evaluation} and it is impractical to report all of them, we systematically selected and reported 50 issues according to the following rules. {(1)} We selected active and popular repositories with high stars in GitHub, which had discussions or active issues within two months before we reported the issues. {(2)} We only selected issues that could be fixed by \tool. For instance, it is difficult to resolve issues caused by licenses of two third-party components unless at least one of the components are replaced by others. {(3)} The selected issues covered most popular licenses such as {the MIT License}, {Zope Public License 2.1}, {Apache License 2.0}, {GPL 3.0}, and {BSD 3-Clause License}. The size of selected projects ranged from 100K to 1G. For each issue, we reported the licenses involved in the incompatibility issues, their locations, the inherent reason for license incompatibility, and the suggestions to resolve this issue. So far, 14 reported issues have been responded as shown in \Cref{tab:FeedBack}. Although the remaining issues have been confirmed by ourselves, they are still waiting for responses from OSS developers.}

\begin{table}
\centering
 \caption{\revise{Feedback from OSS Developers}}
 %\footnotesize
  \label{tab:FeedBack}
\scalebox{0.58}{
\begin{tabular}{lccccc}
\toprule
\textbf{Project} &\textbf{\#Stars} & \textbf{Parent License} & \textbf{Child License} & \textbf{Issue State} \\ \midrule
cookiecutter & 10.2k & BSD-3-clause (/LICENSE) & GPL3.0 (/project\_slug/LICENSE) &{{Ignored}} \\ \hline
\multirow{2}{*}{bypy} & \multirow{2}{*}{6.7k} & MIT (/LICENSE) & GPL3.0 (bypy/gui.py)  &\textcolor{springgreen}{\textbf{Fixed}} \\
& & MIT (/LICENSE) & GPL3.0 (bypy/\_\_init\_\_.py)  &\textcolor{springgreen}{\textbf{Fixed}} \\\hline
%peewee & 9.8k & 2667 & tests/libs/mock.py &{Ignored} \\ \hline
hypnotix & 798 & GPL3.0 (/README.md) & AGPL3.0 (usr/lib/hypnotix/mpv.py) &{{Denied}} \\ \hline
\multirow{3}{*}{mitmproxy} & \multirow{3}{*}{30.1k} & \multirow{3}{*}{MIT (/LICENSE)} & LGPL2.1 (../../raw\_display.py) &\textcolor{springgreen}{\textbf{Fixed}} \\ 
 &  &  & Apache2.0 (../tornado/\_\_init\_\_.py) &\textcolor{springgreen}{\textbf{Fixed}} \\ 
  &  &  & BSD-3-clause (../click/\_\_init\_\_.py) &\textcolor{springgreen}{\textbf{Fixed}} \\ \hline
webssh & 3.4k & MIT (/LICENSE) & LGPL2.1 (tests/sshserver.py) &\textcolor{springgreen}{\textbf{Fixed}} \\ \hline
\multirow{4}{*}{OpenNRE} & \multirow{4}{*}{3.9k} & \multirow{4}{*}{MIT (/LICENSE)} & Apache2.0 (../word\_piece\_tokenizer.py) &\textcolor{springgreen}{\textbf{Confirmed}} \\ 
 &  &  & Apache2.0 (../bert\_tokenizer.py) &\textcolor{springgreen}{\textbf{Confirmed}} \\ 
 &  &  & Apache2.0 (../word\_tokenizer.py) &\textcolor{springgreen}{\textbf{Confirmed}} \\ 
 &  &  & Apache2.0 (../basic\_tokenizer.py) &\textcolor{springgreen}{\textbf{Confirmed}} \\ \hline
 \multirow{2}{*}{websockets} & \multirow{2}{*}{879} & \multirow{2}{*}{MIT (/LICENSE.txt)} & Apache2.0 (ws4redis/websocket.py) &{{Denied}} \\ 
 &  &  & Apache2.0 (ws4redis/utf8validator.py) &{{Denied}} \\ 
 
\bottomrule
\end{tabular}}
\end{table}

\revise{As shown in~\Cref{tab:FeedBack}, 10/14 responses confirmed or fixed the reported issues. One developer ignored the reported issue and 3/14 responses denied that there existed license incompatibility in their projects. Moreover, we also investigate the reasons that prevent some developers from fixing license incompatibility issues and obtain several findings. \textbf{(1)} OSS developers have different awareness and knowledge about OSS licenses. For example, one developer believed that the MIT license and the Apache 2.0 license were two similar licenses and thus should be compatible with each other. Actually, the Apache 2.0 license is more restrictive than the MIT license in several aspects such as \textit{Use Trademarks} and \textit{State Changes}. As a result, when the project license is the MIT license and its child license is the Apache 2.0 license, there exists license incompatibility which might bring legal risks to the users of this project. \textbf{(2)} It is inconvenient for some large-scale projects with many contributors and users to fix license incompatibility issues, since every contributor should be contacted for the agreement to change the licenses which may also affect many users. Therefore, paying attention to the license incompatibility problem especially at an early stage of software development is necessary. \textbf{(3)} There exist different opinions towards the licenses of imported packages. Some developers believed that there was no need to consider the licenses of imported packages, which was actually against the intentions of some licenses (e.g., the Apache 2.0 license, the GPL 3.0 license, and the MPL 2.0 license)}.

\subsection{Lessons Learned}
We summarize several lessons that might be learned from this work. \textbf{First}, developers should pay more attention to the licenses when reusing OSS code. One reason is that the incorporated licenses might be more restrictive than the project license, and thus developers have to modify the regulations for the whole project to comply with the incorporated licenses. The second reason is that licenses accompanied with different third-party software packages might be incompatible with each other. In this case, developers are enforced to either replace the incompatible package with an alternative one, which requires substantial efforts due to the complex package dependency, or attach an exception to the project license that states different regulations towards different source code, which might not be encouraged by the community.
\textbf{Second}, not all license incompatibility issues can be resolved. Among the investigated 230 incompatible projects, {37.39\%} of them cannot be resolved unless developers replace the involved third-party packages. The main reason is that a project owner has no rights to change the licenses accompanied with the integrated third-party packages. Another reason is that the constraints from the license hierarchy cannot be resolved in some cases. For instance, given a license, when its parent license is more permissive than one of its child licenses, the constraints from the parent and child nodes cannot be resolved.
\textbf{Third}, official licenses can only resolve a small part of license incompatibility issues. It can be seen that only {34.52\%} of the incompatibility issues can be addressed by recommending a new official license. Sometimes developers have to create their own licenses (i.e., custom licenses) to satisfy the requirements of all incorporated OSS licenses. Otherwise, they have to migrate the packages whose licenses are involved in incompatibility issues.
\textbf{Finally}, although it is a common practice for developers of large-scale OSS to place their own licenses in the modules implemented by themselves, developers should be aware of the license hierarchy that represents the licensing scope of each license, where a module license should not be more permissive than all the licenses in this module, and should not be more restrictive than the licenses in the upper-level module that contains this module simultaneously.

\subsection{Limitations and Threats to Validity}
The limitations are from two aspects. %\textbf{First}, in the phase of license understanding, we construct two probabilistic models to extract fine-grained entities and their relations based on sequence labelling and prompts, respectively. For this reason, we cannot theoretically ensure that all entities and relations can be extracted by \tool. However, it can be seen that the performance of \tool is superior to that of four baselines in license understanding. 
\textbf{First}, given an arbitrary project, if it has no project license or its project license does not state its copyright holder, \tool cannot find any internal licenses (also known as component licenses) that can be changed according to the copyright holder. %In this case, \tool only assumes that the project license can be modified or replaced by the project owner.
\textbf{Second}, \tool has not taken into consideration irrelevant licenses (e.g., licenses only for testing purposes). Besides, multiple project licenses in the root directory of the project {are considered as the supplements of each other}, which might also affect the performance of \tool on license incompatibility localization and resolution. 

%\subsection{Threats to Validity}
Threats that may affect the results of evaluation include the following two aspects. \textbf{First}, the quality of manual labelling could be a threat to validity. The performance of \tool on license understanding and incompatibility resolution was evaluated on the ground-truth datasets comprising {21,844} license sentences and {300} real-world OSS, the labelling quality of which may
threaten the results. To mitigate this problem, three authors and a lawyer manually verified and cross-validated the datasets. %We made the datasets publicly available for further studies~\cite{ourwebsite}. 
\textbf{Second}, as previous studies~\cite{tldrlegal,xu2021lidetector}, we categorized \textit{actions} into 23 types. However, there could be a few cases where special requirements outside the scope of these actions are claimed by the licensor, which might influence the results of \tool.
%
%\textbf{Third}, for recommending official licenses, although \tool can automatically understands the regulations stated by official licenses, \tool utilizes a database constructed in the preparation phase to find the ones that satisfy the constraints for incompatibility resolution. Nevertheless, there might be some official licenses not collected by the database, and thus will be omit by \tool when recommending existing licenses. To mitigate this issue, more than {200} widely-used official licenses are considered when recommending licenses, and the database can be further improved in the future work. 

\section{Related Work}
\subsection{License Detection and Semantic Extraction}
%To facilitate the process of understanding and choosing appropriate licenses, m
Much research has been done to automatically identify official licenses and extract the semantics from license texts. 

\noindent \textbf{License detection}. %Many studies have been conducted to identify the versions and names of some commonly-used licenses. 
Gobeille et al.~\cite{msr`08-Gobeille-fossology} presented the first study on license detection. They exploited a binary Symbol Alignment Matrix algorithm to identify OSS licenses.
Tuunanen et al.~\cite{ase`09-Timo-ASLA} designed license templates and identified some well-known licenses based on regular expressions. %They considered both the license texts at the beginning of each source code file and the licenses referenced from the source code. %Similarly, Xu et al.~\cite{cise`10-Xu-OSLC} utilized regular matching to identify licenses including their names and versions. 
Ninka~\cite{ase`10-German-Ninka}, a notable tool for automated license identification, was implemented based on sentence-matching. %It identified licenses through four types of prior knowledge, i.e., filtering keywords, equivalence phrases, known sentence-token expressions, and  a set of license rules. 
%Ninka covers 112 popular OSS licenses, but its recognition range is relatively limited due to the increasing new or custom licenses.
%that continue to appear today. 
On this basis, Higashi et al.~\cite{IWESEP`16-Higashi-clusteringForNinka} employed a clustering algorithm to further identify licenses that could not be handled by Ninka. %Despite being time-efficient, these studies are limited to certain licenses and rely on prior knowledge from experts. This paper enlarges the range of licenses that can be handled automatically.

\noindent \textbf{Semantic extraction}. 
%License texts are typically long and complicated, which motivated some studies on automatic semantic extraction. 
To extract license semantics, many studies conducted the ontology study on licenses~\cite{IREC`09-Alspaugh-Intellectual,AIS`10-Alspaugh-Challenge,Qualipso`10-gordon-prototype,ICAIL`11-Gordon-Analyzing,ICAIL`13-Gordon-Introducing,COMMA`14-Gordon-Demonstration}. For example, Alspaugh et al.~\cite{IREC`09-Alspaugh-Intellectual}~\cite{AIS`10-Alspaugh-Challenge} extracted tuples from licenses to model 10 licenses. 
%Gordon et al.~\cite{Qualipso`10-gordon-prototype} used the Web Ontology Language to build the ontology for 8 OSS licenses. 
Unlike the ontology-based approaches that required much prior knowledge, Kapitsaki et al.~\cite{APSEC`17-Kapitsaki-termsIdentifying} proposed FOSS-LTE to identify license terms with a topic model. %\revise{Specifically}, they first mapped licenses terms with sentences, and then mapped sentences with topics via Latent Dirichlet Allocation (LDA). %By this means, it built the relations between terms and topics. 
Despite the progress, the topic model might induce much noise and cause low accuracy~\cite{xu2021lidetector}.

\subsection{License Incompatibility Detection}
%To avoid legal and financial risks induced by copyright infringement, much works has been done to detect license incompatibility. 
A major of studies on license incompatibility detection are the graph-based approaches~\cite{jss`17-Kapitsaki-SPDX,SRDSCB`14-Kapitsaki-SPDX,Wheeler`07-Wheeler,paschalides2016validate}. 
%such as Kapitsaki et al.~\citep{jss`17-Kapitsaki-SPDX}~\citep{SEKE`12-Kapitsaki-SPDX}~\citep{SRDSCB`14-Kapitsaki-SPDX} and Wheeler’s~\citep{Wheeler`07-Wheeler}.
Generally, these studies manually constructed a directed graph to represent the compatibility relationships between licenses. %, where a node represents a license and its version, and an edge represents the transitive relation between two adjacent licenses. %The directionality in the graph describes the ``one-way compatbility'' relations between licenses, i.e., compatibility stems from a less restrictive license to amore restrictive one. 
Then, they detected license incompatibility issues by examining the graph to determine whether two licenses can reach the same node. %, which means these licenses can be combined in a new software product under an alternative license. %{For instance, Kapitsaki et al.~\cite{paschalides2016validate} proposed a tool named SLVT, which is based on the directed acyclic license graph, to examine license violations that may exist in a single or multiple SPDX files.} 
Despite being strict, %the performance of these methods depends on the presence of target licenses in a hand-crafted graph. However, 
there are only a small number of licenses are supported (e.g., 20 licenses and their versions supported by SPDX-VT~\cite{jss`17-Kapitsaki-SPDX}). %Actually, there are various licenses in the real-word OSS, including many commonly-used licenses, their versions, exceptions, and even custom licenses. Hence, i
It is difficult to manually analyze all the compatibility relations between licenses. Unlike these approaches, \tool provides a flexible solution that can be applied for arbitrary licenses without prior knowledge. %, which makes it more feasible for detecting and fixing license incompatibility issues.

LiDetector~\cite{xu2021lidetector} was the first work that proposed a machine learning-based method for license incompatibility detection. \revise{The main differences between \tool and LiDetector range from license understanding to incompatibility detection. \textbf{(1)} Different granularities of entities. LiDetector regarded each right/obligation as a single entity, while \tool splits it into four types of entities, followed by a relation extraction model to organize them. By this means, \tool models a regulation more accurately, which as a result benefits license incompatibility detection and resolution. \textbf{(2)} Different models. \tool first embeds words with a roberta-base model, and then feeds them into a transition-based parser. In addition, LiDetector regarded each entity independently, while \tool employs a prompt-based model to extract the relations between entities. \textbf{(3)} Different licensing scopes. LiDetector treated all licenses except the project license equally and independently, while \tool observes that the location of a license determines its scope of licensing, and thus extracts license hierarchy for license incompatibility detection and resolution. \textbf{(4)} Different detection strategies. Based on license hierarchy extraction, \tool obtains \textit{parent-child} pairs of licenses from the license hierarchy, and defines that a license is compatible with its child nodes if {anyone who complies with the license will not violate any of its child nodes}. On this basis, \tool detects license incompatibility from bottom to up, layer by layer according to license hierarchy. Moreover, the detection strategy of \tool is more fine-grained than LiDetector, which benefits from its fine-grained license understanding.}

\subsection{License Recommendation}
%Another aspect is license recommendation or prediction.
%The findOSSLicense tool~\citep{tse`18-Kapitsaki-findOSSLicense} of Kapitsaki et al. 
Based on the aforementioned graph-based methods for license incompatibility detection, Kapitsaki et al.~\cite{kapitsaki2019modeling} proposed FindOSSLicense to recommend OSS licenses. It considered user requirements through a set of answers to questions, the licenses used by  similar users or similar projects, and also license compatibility information according to the directed graph proposed in~\cite{jss`17-Kapitsaki-SPDX}. To this end, a hybrid approach was proposed to combine the content-based, constraint-based, and
collaborative filtering techniques in the recommendation system. 
%Although FindOSSLicense took license incompatibility into consideration when recommending licenses, 
However, the license graph only covered a small number of licenses. Licenses not covered in the license graph were marked with ``caution''. %Moreover, it did not take into account licenses accompanied with third-party components which are not explicitly declared in the OSS. 
Liu et al.~\cite{liu2019predicting} observed that software changes might lead to license updates. Based on this observation, they proposed to predict source code
file-level licenses in the
presence of software changes. % Specifically, they collected a large corpus of changed files with their licenses and leveraged machine learning techniques to predict license conflicts and the resolutions. 
However, the license prediction tool is only applicable for software changes, which limits its application scope. %Moreover, it only considered 25 licenses in the prediction models, leading to the performance degradation when faced with other licenses.
Compared with these approaches,
\tool automatically models license texts in a fine-grained way, analyzes the inherent reasons of license incompatibility for arbitrary licenses, and provides flexible solutions for resolving license incompatibility issues. %such as recommending a commonly-used license and generating custom licenses for users to choose.} 
%the method in this paper has realized the automatic extraction of the license terms and their authorization types. Therefore, we can automatically analyze the compatibility relationship among the licenses in a OSS project and are not restricted by the scope of the license, which has wider application scenarios and higher effectiveness. 
\section{Conclusion}
In this paper, we propose \tool, an {automated} tool to resolve license incompatibility issues for open source software. Given an OSS, it first extracts all licenses along with their licensing scopes and dependencies. 
Then, it conducts a fine-grained understanding of license texts, based on which it detects and localizes license incompatibility issues. Finally, for each incompatible license that can be modified by the software owner, 
it computes the constraints from the context of the license, and provides useful and flexible suggestions. Comprehensive experiments on 300 real-world OSS demonstrate the effectiveness of \tool. %, with {4.09\%} FP rate and {0.02\%} FN rate for incompatibility issue localization, and {62.61\%} success rate for resolving 230 incompatible OSS on GitHub. %Moreover, we also evaluate the impacts of license hierarchy and copyright holder on incompatibility resolution. 
%We made all datasets and the replication package of \tool publicly available to facilitate follow-up research.

\section*{Acknowledgements}
%We thank the anonymous reviewers for their valuable suggestions. 
This work was supported by the National Key Project of China (No. 2020YFB1005700), the National Natural Science Foundation of China (No. 62202245, 62102197, and 62002178), and the National Science Foundation of Tianjin (No. 22JCYBJC01010).

%\clearpage

\balance

\bibliographystyle{ACM-Reference-Format}
\bibliography{sample-base}


\balance
\end{document}